\documentclass[12pt]{iopart}
\usepackage{iopams}
\usepackage{setstack}
\usepackage{cancel} 
\usepackage{enumerate}
\usepackage{graphicx}
\usepackage{epsfig} 
\usepackage{subfig}
\usepackage{caption}
\usepackage{booktabs}
\usepackage{color}
\usepackage{hyperref}

\begin{document}
\title{On the black hole limit of rotating discs of charged dust}
\author{Martin Breithaupt, Yu--Chun Liu, Reinhard Meinel and Stefan Palenta}
\address{Theoretisch--Physikalisches Institut, Friedrich Schiller Universit\"at Jena, Max--Wien--Platz 1, 07743 Jena, Germany}
\eads{\mailto{martin.breithaupt@uni-jena.de}, \mailto{yu-chun.liu@uni-jena.de}, \mailto{meinel@tpi.uni-jena.de}, \mailto{stefan.palenta@uni-jena.de}}

\begin{abstract}
Investigating the rigidly rotating disc of dust with constant specific charge, we find that it leads to an extreme Kerr--Newman black hole in the ultra--relativistic limit. A necessary and sufficient condition for a black hole limit is, that the electric potential in the co--rotating frame is constant on the disc. In that case certain other relations follow. These relations are reviewed with a highly accurate post--Newtonian expansion.

Remarkably it is possible to survey the leading order behaviour close to the black hole limit with the post--Newtonian expansion. We find that the disc solution close to that limit can be approximated very well by a ``hyper\-extreme'' Kerr--Newman solution with the same gravitational mass, angular momentum and charge.
\end{abstract}
\vspace{2pc}

\submitto{\CQG}
 
\section{Introduction}
Bardeen and Wagoner \cite{BardWag} were able to solve the general--relativistic problem of a uniformly rotating disc of dust
by means of a post--Newtonian expansion to very high order. Although their method involved the successive numerical evaluation of integrals
they achieved an impressive accuracy even in the extreme relativistic limit. From the exterior perspective, the spacetime approaches the metric of an extreme Kerr black hole (outside the horizon) in that limit. This remarkable prediction was rigorously confirmed with the exact solution to the disc problem by Neugebauer and Meinel \cite{nm95}, see also \cite{m02}. Starting from the exact solution, Petroff and Meinel \cite{pm01} presented an analytic post--Newtonian expansion scheme up to arbitrary order. Recently \cite{MeinPal}, a similar analytic expansion scheme was found for the more general problem of a uniformly rotating disc of electrically charged dust (with a constant specific charge) -- without knowing the corresponding exact solution. Again, evidence for a black hole limit was given. Based on a further elaboration of the method presented in \cite{MeinPal}, the aim of the present paper is a detailed investigation of the electrically charged disc of dust in the extreme relativistic limit. In particular, the behaviour of the gravitational and electromagnetic multipole moments is studied. The results are fully consistent with a spacetime that approaches, from the exterior perspective, the metric of an extreme Kerr--Newman black hole in this limit.
\par
The paper is organised as follows. Section \ref{sec:Desc_CD} is an introduction to rotating charged dust and its mathematical description. Here we also give a proof for an important parameter equation. In section \ref{sec:PN_cDisc} we give an outline to the post--Newtonian expansion of rigidly rotating discs with constant specific charge and how we can treat the question of its convergence. Section \ref{sec:bhl} is dedicated to the black hole limit of these discs, and in section \ref{sec:LOB_BL} we make statements on the leading order behaviour close to that limit. In the Appendix we have provided the first few analytic coefficients of the expansion for the relevant quantities within this paper.
 
\section{Description of rotating configurations of charged dust in equilibrium} \label{sec:Desc_CD}
\subsection{Basic field equations and model of matter}
The Einstein--Maxwell equations read\footnote{We use geometrized Gauss units ${\rm c}={\rm G}=4\pi\epsilon_0=1$ and the metric signature $(+,+,+,-)$. Covariant differentiation is denoted by a semicolon and partial differentiation is denoted by a comma.}
\begin{eqnarray}
 & R_{ab} - \frac{1}{2}Rg_{ab} = 8\pi T_{ab}, \label{EFG}\\
 & F_{[ab;c]}=0, \label{hom_maxwell_eq}\\
 &  F^{ab}_{\quad;b}=4\pi \jmath^a. \label{inhom_maxwell_eq}
\end{eqnarray}
The homogeneous Maxwell equations (\ref{hom_maxwell_eq}) are fulfilled by a four--potential $A_a$ with
\begin{eqnarray}
F_{ab} = A_{b;a}-A_{a;b} = A_{b,a}-A_{a,b}.
\end{eqnarray}
The energy momentum tensor $T_{ab}$ for the model of charged dust that we use can be expressed as the sum of a dust part and an electromagnetic part
\begin{eqnarray}
 T_{ab} = T^{{(\rm dust})}_{ab} + T^{({\rm em})}_{ab} = \mu u_au_b + \frac{1}{4\pi}\left(F_{ac}F_b^{\;c}-\frac{1}{4}g_{ab}F_{cd}F^{cd}\right). \label{EMT_dust}
\end{eqnarray}
The baryonic mass density $\mu$ is related to the charge density $\rho_{\rm el}$ via
\begin{eqnarray}
 \rho_{\rm el} = \epsilon\mu,
\end{eqnarray}
where in general the specific charge $\epsilon$ is a free function. We assume a purely convective four--current density, i.e. 
\begin{eqnarray}
 \jmath^a = \rho_{\rm el} u^a = \epsilon\mu u^a.
\end{eqnarray}
\par
Now we consider isolated and rotating configurations of charged dust in equilibrium, which leads to a stationary spacetime, giving a Killing vector $\bxi$. Furthermore, assuming axisymmetry, we have a second Killing vector $\bfeta$ that commutes with $\bxi$. For an asymptotically flat spacetime, the far field behaviour of the Killing vectors is given by
\begin{eqnarray}
 \xi^a\xi_a = - 1 + \frac{2M}{r} + {\cal O}\left(r^{-2}\right), \quad \frac{\xi^a\eta_a}{\eta^b\eta_b} = - \frac{2J}{r^3} + {\cal O}\left(r^{-4}\right), \label{farfieldMJ}
\end{eqnarray}
with $r=\sqrt{x^2+y^2+z^2}$ and $x,y,z$ being asymptotic Cartesian coordinates, the gravitational mass $M$ and the angular momentum $J$. The metric can be written globally in terms of Lewis--Papapetrou--Weyl coordinates
\begin{eqnarray}
\mathrm{d}s^2=f^{-1}\left[h\left(\mathrm{d}\varrho^2+\mathrm{d}\zeta^2\right)+\varrho^2\mathrm{d}\varphi^2\right]-f\left(\mathrm{d}t+a\mathrm{d}\varphi\right)^2, \label{Weylmetrik}
\end{eqnarray}
with
\begin{eqnarray}
 \bxi = \partial_t, \quad \bfeta = \partial_\varphi.
\end{eqnarray}
For a more detailed discussion of equilibrium states and the far field see \cite{RelFig}. Using Lorenz gauge, we get $A_a = (0,0,A_\varphi,A_t)$.
\par
Altogether, we have to solve the Einstein--Maxwell equations for the five unknown functions $f,h,a, A_t$ and $A_\varphi$ which depend only on the coordinates $\varrho$ and $\zeta$ while the angular velocity $\Omega=\frac{\rmd\varphi}{\rmd t}$ and the specific charge $\epsilon$ are specified by a particular physical model. 
\par
Here we choose rigid rotation ($\Omega=$ constant) and a constant specific charge $\epsilon$. In our units we have $\epsilon\in[-1,1]$, obtaining static electrically counterpoised dust (ECD) con\-figurations (see \cite{BW71} and \cite{Meinel2011}) for $\epsilon=\pm1$. In the case of rigid rotation, a global co--rotating frame exists with $\varphi'=\varphi-\Omega t$, where the metric retains its form
\begin{eqnarray}
\mathrm{d}s'^2=f'^{-1}\left[h'\left(\mathrm{d}\varrho'^2+\mathrm{d}\zeta'^2\right)+\varrho'^2\mathrm{d}\varphi'^2\right]-f'\left(\mathrm{d}t'+a'\mathrm{d}\varphi'\right)^2 \label{Weylmetrik_prime}
\end{eqnarray}
and the metric functions and electromagnetic potentials transform as
\begin{eqnarray} 
 &f' = f\left(1+\Omega a\right)^2-\frac{\Omega^2\varrho^2}{f}, \quad \left(1+\Omega a\right)f = \left(1-\Omega a'\right)f', \quad \frac{h}{f}=\frac{h'}{f'} \\
 &A'_{t'} = A_t+\Omega A_\varphi, \quad A'_{\varphi'} = A_\varphi.
\end{eqnarray}
The four velocity in the co--rotating frame can be expressed as
\begin{eqnarray}
 u'^{n} = \left(0,0,0,1/\sqrt{f'}\right), \quad u'_{n} = \left(0,0,-a'\sqrt{f'},-\sqrt{f'}\right). \label{u_prime}
\end{eqnarray}

\subsection{Ernst equations}
Outside the dust configuration, we apply the Ernst equations \cite{Ernst} which are equivalent to the stationary and axisymmetry electro--vacuum Einstein--Maxwell equations. Using the abbreviation 
\begin{eqnarray}
 A_t=-\alpha,
\end{eqnarray}
they read\footnote{We use the operators $\nabla$ and $\Delta$ as in Euclidean 3--space, with cylindrical coordinates ($\varrho,\zeta,\varphi$).} 
\begin{eqnarray}
  \left(\Re\mathcal{E}+\bar{\Phi}\Phi\right)\Delta\mathcal{E} =&\; \left(\nabla\mathcal{E}+2\bar{\Phi}\nabla\Phi\right)\cdot\nabla\mathcal{E}, \label{Ernst_E}\\
 \left(\Re\mathcal{E}+\bar{\Phi}\Phi\right)\Delta\Phi =&\; \left(\nabla\mathcal{E}+2\bar{\Phi}\nabla\Phi\right)\cdot\nabla\Phi \label{Ernst_Phi}
\end{eqnarray}
with
\begin{eqnarray}
 {\cal E} = f - \Phi\bar\Phi + {\rm i}b, \quad \Phi = \alpha + {\rm i}\beta
\end{eqnarray}
and
\begin{eqnarray}
 &b_{,\varrho}     = -\frac{f^2}{\varrho}a_{,\zeta} - 2\alpha\beta_{,\varrho} + 2\beta\alpha_{,\varrho}, \quad b_{,\zeta} = \frac{f^2}{\varrho}a_{,\varrho} - 2\alpha\beta_{,\zeta} + 2\beta\alpha_{,\zeta}, \\
 &\beta_{,\varrho} = \frac{f}{\varrho}\left(a\alpha_{,\zeta}+A_{\varphi,\zeta}\right), \quad \beta_{,\zeta} = -\frac{f}{\varrho}\left(a\alpha_{,\varrho}+A_{\varphi,\varrho}\right).
\end{eqnarray}
Here the differential equations for the metric function $h$ are decoupled from the others. It can be calculated via a path--independent line integral afterwards:
\begin{eqnarray}
\fl \left(\ln h\right)_{,\varrho} = \frac{1}{2}\varrho\left(\left(\ln f\right)_{,\varrho}^2-\left(\ln f\right)_{,\zeta}^2 - \frac{f^2}{\varrho^2}\left(a_{,\varrho}^2-a_{,\zeta}^2\right)\right) +2\frac{f}{\varrho}\left(A^2_{\varphi,\varrho}-A^2_{\varphi,\zeta}\right) \nonumber\\ -\, 2\frac{\varrho^2-a^2f^2}{f\varrho}\left(A^2_{t,\varrho}-A^2_{t,\zeta}\right) -4\frac{af}{\varrho}\left(A_{\varphi,\varrho}A_{t,\varrho}-A_{\varphi,\zeta}A_{t,\zeta}\right), \\
\fl \left(\ln h\right)_{,\zeta} = \varrho\left(\left(\ln f\right)_{,\varrho}\left(\ln f\right)_{,\zeta} - \frac{f^2}{\varrho^2}a_{,\varrho}a_{,\zeta}\right) +4\frac{f}{\varrho}A_{\varphi,\varrho}A_{\varphi,\zeta}-4\frac{\varrho^2-a^2f^2}{f\varrho}A_{t,\varrho}A_{t,\zeta} \nonumber\\ -\,4\frac{af}{\varrho}\left(A_{\varphi,\zeta}A_{t,\varrho}+A_{\varphi,\varrho}A_{t,\zeta}\right).
\end{eqnarray}

\subsection{Multipole moments}
We introduce the potentials 
\begin{eqnarray}
 \Xi = \frac{1-{\cal E}}{1+{\cal E}} \quad{\rm and}\quad q = \frac{2\Phi}{1+{\cal E}}.
\end{eqnarray}
On the positive axis of symmetry they can be written as a series expansion in $\zeta^{-1}$:
\begin{eqnarray} \label{def_mn_en}
 \Xi_+ = \frac{1}{\zeta}\sum\limits_{n=0}^{\infty}\frac{m_n}{\zeta^{\,n}}, \quad  q_+ = \frac{1}{\zeta}\sum\limits_{n=0}^{\infty}\frac{e_n}{\zeta^{\,n}}.
\end{eqnarray}
The gravitational multipole moments $P_n$ and the electro--magnetic multipole moments $Q_n$ introduced by Simon \cite{Simon} can be obtained from the coefficients $m_n$ and $e_n$ (see \cite{HP} and \cite{SA}). For $n\leq3$ they simply read $P_n=m_n$ and $Q_n=e_n$.

\subsection{Physical quantities}
In \cite{Proceed} we presented the parameter equation
\begin{eqnarray}
 M = 2\Omega J + {\cal D}M_0
\end{eqnarray}
for the gravitational mass. Here we will give a short derivation. The Lorentz force density is given by
\begin{eqnarray}
 {\tt f}^a = -\left(T^{({\rm em})ab}\right)_{;b}=F^{ab}\jmath _b
\end{eqnarray}
and orthogonal to the Killing vectors
\begin{eqnarray}
 -\xi_a{\tt f}^a = \left(\xi_aT^{({\rm em})ab}\right)_{;b} = 0, \quad -\eta_a{\tt f}^a = \left(\eta_aT^{({\rm em})ab}\right)_{;b} = 0. \label{K_Lfd}
\end{eqnarray}
Gauss law, the local conservation of energy momentum and (\ref{K_Lfd}) give us the option to define a set of physical quantities in a coordinate independent way.
\par
Gravitational mass $M$ and angular momentum $J$ are given by (see also \cite{Wald})
\begin{eqnarray}
 M = 2\int\limits_{V}\left(T_{ab}+\frac{1}{2}\mu g_{ab}\right)\xi^an^b{\rm d}{\cal V}, \quad J = -\int\limits_{V}T_{ab}\eta^an^b{\rm d}{\cal V} \label{Def_M_J}
\end{eqnarray}
where $V$ is a space--like hyper--surface with the 3--dimensional volume element $\rmd{\cal V}=\sqrt{^{(3)}g}\,\rmd^3x$ and the future pointing unit normal $n^a$. The definitions of the gravitational mass $M$ and the angular momentum $J$ are arranged in such a way, that they coincide with the far field behaviour in (\ref{farfieldMJ}). From the local mass conservation law $\left(\mu u^a\right)_{;a}=0$ and the continuity equation $\jmath^a_{\;\,;a}=0$, we get the baryonic mass $M_0$ and the total charge $Q$ as
\begin{eqnarray}
 M_0 = {\int\limits_{V_{\rm dust}}}{\rm d}M_0 = -{\int\limits_{V_{\rm dust}}}\mu u^an_a{\rm d}{\cal V}, \quad  Q = {\int\limits_{V_{\rm dust}}}{\rm d}Q = -{\int\limits_{V_{\rm dust}}}\epsilon\mu u^an_a{\rm d}{\cal V}. \label{Def_M0_Q}
\end{eqnarray}
With (\ref{K_Lfd}) we can define the electromagnetic field energy $M_{\rm em}$ and the electromagnetic field angular momentum $J_{\rm em}$ as
\begin{eqnarray}
 M_{\rm em} = \int\limits_{V}T^{({\rm em})}_{ab}\xi^an^b{\rm d}{\cal V}, \quad J_{\rm em} = -\int\limits_{V}T^{({\rm em})}_{ab}\eta^an^b{\rm d}{\cal V}. \label{Def_Mem_Jem}
\end{eqnarray}
The definitions in (\ref{Def_Mem_Jem}) are arranged in a way, that they coincide with the electromagnetic field energy and the electromagnetic field angular momentum in classical electro\-dynamics.
\par
In Lewis--Papapetrou--Weyl coordinates we integrate (\ref{Def_Mem_Jem}) by parts and use the inhomogeneous Maxwell equations (\ref{inhom_maxwell_eq}). This leads to
\begin{eqnarray}
 M_{\rm em} = \frac{1}{2}\int\limits_{V_{\rm dust}}A_{a}\jmath _b\left(\xi^a-\Omega\eta^a\right)n^b\,{\rm d}{\cal V}, \quad J_{\rm em} = -\int\limits_{V_{\rm dust}}A_{a}\jmath _b\,\eta^an^b\,{\rm d}{\cal V}. \label{Def_Mem_Jem_dust}
\end{eqnarray}
So the domain of integration $V$ in (\ref{Def_Mem_Jem}) (and therefore also in (\ref{Def_M_J})) can be reduced to the volume of the dust. For rigid rotation we can combine (\ref{Def_Mem_Jem_dust}) and (\ref{Def_M_J}) with the help of (\ref{u_prime}) to
\begin{eqnarray}
 M - 2\Omega J = {\int\limits_{V_{\rm dust}}}\left(\sqrt{f'}+\epsilon\alpha'\right){\rm d}M_0
\end{eqnarray}
where $f'$ and the electric potential $\alpha'=-A'_{t'}$ refer to the co--rotating system. For a constant specific charge, it follows from the equations of motion 
\begin{eqnarray} \label{EoM}
\mu\dot{u}^a = {\tt f}^a,
\end{eqnarray} 
evaluated in the co--rotating frame, that 
\begin{eqnarray} \label{Dconstant}
\sqrt{f'}+\epsilon\alpha' = {\cal D} = {\rm constant}
\end{eqnarray}
inside the dust and so we get
\begin{eqnarray} \label{M_formula}
M = 2\Omega J + {\cal D}M_0
\end{eqnarray}
independent from the shape of the dust configuration.
 
\section{Post--Newtonian expansion of rigidly rotating discs with constant specific charge} \label{sec:PN_cDisc}
\subsection{Mathematical problem}
The dust configuration shall be a circular disc with the coordinate radius $\varrho_0$, centred in the origin of the Lewis--Papapetrou--Weyl coordinate system ($\zeta=0, 0\leq\varrho\leq\varrho_0$). The mass density $\mu$ can be written formally as $\mu=\sigma_{\mathrm p}(\varrho)\sqrt{f/h}\,\delta(\zeta)$ with the proper surface mass density $\sigma_{\mathrm p}(\varrho)$. By integrating the full Einstein--Maxwell equations over a small flat cylinder around a mass element of the disc, matching conditions between the metric functions and electromagnetic potentials (and their derivatives) at $\zeta=0^{\pm}$ (above and below the disc) can be derived. The interior geometry of the disc itself and the surface mass density can be calculated from these functions as well. Assuming reflectional symmetry, we showed in \cite{MeinPal} that the whole problem can thus be reformulated as a boundary value problem to the Ernst equations (\ref{Ernst_E}) and (\ref{Ernst_Phi}) with the boundary conditions
\begin{eqnarray} \label{BC}
\Im{\cal E'}=0, \;  \Im{\Phi'}=0, \; \varrho_0\left(\Re\Phi'+\epsilon\sqrt{f'}\right)_{,\zeta}=0, \; \varrho_0\left(\sqrt{f'}+\epsilon\Re\Phi'\right)_{,\varrho}=0
\end{eqnarray}
on the disc. The last one is equivalent to (\ref{Dconstant}). We also presented a powerful post--Newtonian expansion in the parameter
\begin{eqnarray}
 \gamma=1-\sqrt{f_{\rm c}} \quad{\rm where}\quad f_{\rm c} =  f(\varrho=0,\zeta=0)
\end{eqnarray}
and an algorithm to solve the equations to arbitrary order in $\gamma$ analytically as well as some results for the first eight orders. Here we extended the calculations to the tenth order, where the general structure of the solution (global pre--factors, polynomial structure, etc.), described in \cite{MeinPal}, is still valid.

\subsection{New charge parameter}
The disc solution is a three parameter solution $\mathbb{L}(p_{\rm s},p_1,p_2)$, with $p_{\rm s}$ as a scaling parameter. Here we will use the coordinate radius of the disc $\varrho_0$ or the gravitational mass $M$ as scale parameters. Functions and coordinates scaled with $\varrho_0$ are labelled with a $^*$ and those scaled with $M$ are labelled with a $^\circ$. The other two parameters are $\gamma$ (strictly speaking $\sqrt{\gamma}$) and $\epsilon=Q/M_0$. For the purpose detailed in section \ref{sec:bhl}, it is more effective to use the dimensionless parameter $\psi=Q/M$ instead of $\epsilon$, that is given as
\begin{eqnarray}
 \psi(\gamma,\epsilon) = \epsilon\left\{ 1 + \left(1-\epsilon^2\right)\left[\frac{1}{5}\gamma + \frac{1}{175}\left(21-11\epsilon^2\right)\gamma^2 + {\cal O}\left(\gamma^3\right)\right] \right\},
\end{eqnarray}
containing $\epsilon$ as a global pre--factor. In the ECD--limit ($\epsilon=\pm1$) we have $M=M_0$ and therefore $\psi\in[-1,1]$. Functions with a global pre--factor $\sqrt{1-\epsilon^2}$ also have a global pre--factor $\sqrt{1-\psi^2}$.

\subsection{Question of convergence}
A clear statement to the radius of convergence for a power series (\ref{gen_pow_ser}) could only be made, if all coefficients up to infinity, or the full analytic expression are known.
\begin{eqnarray} \label{gen_pow_ser}
 F(\gamma,\epsilon) = F_{\rm NL}(\gamma,\epsilon)\left[1 + \sum\limits_{n=1}^\infty c_n(\epsilon)\gamma^n\right],
\end{eqnarray}
with $F_{\rm NL}(\gamma,\epsilon)$ describing the Newtonian limit. Real calculations could only be done to a finite value of $n$:
\begin{eqnarray} \label{gen_pow_ser_approx}
 F(\gamma,\epsilon,N) = F_{\rm NL}(\gamma,\epsilon)\left[1 + \sum\limits_{n=1}^N c_n(\epsilon)\gamma^n\right].
\end{eqnarray}
Here we know the first nine coefficients $(N=9)$. For $N\geq4$ all terms in the expansion of the Ernst equations (\ref{Ernst_E}) and (\ref{Ernst_Phi})  contribute to the result. So it seems very likely, that the first nine coefficients $c_n(\epsilon)$ indicate the general behaviour and in particular the convergence of the expansion functions from the post--Newtonian expansion.
\par
Nevertheless, we can review the situation from a physical point of view. The expansion parameter is strictly speaking $\sqrt{\gamma}$ (see the equations (34) in \cite{MeinPal}). So we want to know if we have a well defined physical system in the parameter space $\sqrt{\gamma}\in[-1,1]$ and ${\epsilon}\in[-1,1]$. All functions are either even or odd in $\sqrt{\gamma}$ and either even or odd in $\epsilon$. Functions related to the rotation of the disc are odd functions in $\sqrt{\gamma}$ and have the global prefactor $\sqrt{1-\epsilon^2}$. (Note that $\Omega^*_{\rm NL}=\sqrt{\gamma}\sqrt{1-\epsilon^2}$.) Functions related to the charge of the disc are odd functions in $\epsilon$. In the limit $\sqrt{\gamma}\to+1$ we will approach the ultra--relativistic limit. Changing the sign of $\sqrt{\gamma}$ to negative values results in a disc rotating in the opposite direction. As the potentials $b$ and $\beta$ will change their sign, we get the complex conjugated Ernst potentials that satisfy the Ernst equations as well.
\par
We can also describe the parameter transition from positive to negative values of $\sqrt{\gamma}$. For small values of $\sqrt{\gamma}$ we reach the Newtonian limit. In the limit $\sqrt{\gamma}\to0$ the angular velocity $\Omega$ and the mass density $\mu$ will go to zero. The global solution is the Minkowski spacetime. For small negative values of $\sqrt{\gamma}$ we reach the Newtonian limit of the disc with a negative angular velocity $\Omega$. So we have a smooth transition from positive to negative values in $\sqrt{\gamma}$.
\par
While the sign of $\sqrt{\gamma}$ is related to the direction of the rotation of the disc, the sign of $\epsilon$ is related to the sign of its charge.  For $\epsilon=\pm1$ we reach the static ECD--limit and the equations of motion (\ref{EoM}), which can be interpreted as a generalised force balance for every dust particle inside the disc, can now be interpreted as the balance of electric repulsion and gravitation (see \cite{BW71}). For $|\epsilon|>1$ the electric repulsion cannot be compensated, so the spacetime would not be stationary anymore.
\par
In summary, from the physical point of view it is possible for the post--Newtonian expansion to converge for all $\sqrt{\gamma}\leq1$ if $|\epsilon|\leq1$. It is sufficient to inspect only positive values of $\sqrt{\gamma}$ and $\epsilon$.
\par
A sufficient condition for the convergence of the power series (\ref{gen_pow_ser}) for ${\gamma}\in[0,1]$ is the generalised Raabe--Duhamel test: An infinite series $\sum\limits_{n=0}^\infty c_n(\epsilon)$ is absolutely convergent if
\begin{eqnarray} \label{suff_cond}
\mathbb{K}_n = n\left(1-\left|\frac{c_{n+1}(\epsilon)}{c_n(\epsilon)}\right|\right) > 1, \quad \epsilon\in[0,1]
\end{eqnarray}
for all $n\geq n_0$. It was used by Bardeen and Wagoner in \cite{BardWag} describing the uncharged case. There the $c_n$ are simply numbers, while here we have to deal with functions in $\epsilon$ or $\psi$. So (\ref{suff_cond}) is useful in many cases, but it fails if the $c_n(\epsilon)$ have zeros in $\epsilon$ which depend on $n$. This does not necessarily mean that the power series is not converging, cf. Figure \ref{SOm}.
\par
Here we can check the convergence criterion (\ref{suff_cond}) only up to $n=N=9$. So it can be used to support the assumption that $N=9>n_0$ for the most quantities from the post--Newtonian expansion we have studied here. This is because the $\mathbb{K}_n$ are increasing with increasing $n$. In many cases the increasing of the $\mathbb{K}_n$ is most slowly for $\epsilon=1$, which is the limit of stationarity.
\par
As a last support for the convergence of (\ref{gen_pow_ser_approx}) we investigate its Pad\'e approximation, which gives highly accurate results, as compared with the exact solution \cite{nm95} for the uncharged case even in the ultra--relativistic limit, see \cite{pm01}.

\subsection{Pad\'e approximation}
In many cases the convergence of a power series could be accelerated by using the Pad\'e approximation (see \cite{BenOrs}). The Pad\'e approximation to (\ref{gen_pow_ser_approx}) is given by
\begin{eqnarray} \label{Pade_gen_pow_ser_approx}
 P\,[L,2N-L]\left\{F(\gamma,\epsilon,N)\right\} = F_{\rm NL}(\gamma,\epsilon)\left[\frac{1 + \sum\limits_{n=1}^{2N-L} p_n(\epsilon)\sqrt{\gamma}^{\,n}}{1 + \sum\limits_{n=1}^{L} s_n(\epsilon)\sqrt{\gamma}^{\,n}} \right].
\end{eqnarray}
Therein, the coefficient functions $p_n(\epsilon)$ and $s_n(\epsilon)$ could be determined by the $c_n(\epsilon)$. For that purpose we can equate the coefficients in $\sqrt{\gamma}$ from the series expansion of (\ref{Pade_gen_pow_ser_approx}) with (\ref{gen_pow_ser_approx}). For the Pad\'e approximation we use $\sqrt{\gamma}$ to have a wider scope in choosing $L$ to avoid or minimize the numbers of zeros in the denominator of (\ref{Pade_gen_pow_ser_approx}).

\section{The black hole limit} \label{sec:bhl}

Here we will collect results from the post--Newtonian expansion that indicate most strongly  
the transition to an extreme Kerr--Newman black hole at ${\gamma\to1}$. 
\par
It was shown in \cite{Meinel2004, Meinel2006} that a necessary and sufficient condition for a black hole limit 
of a uniformly rotating (uncharged) fluid body is given by $M-2\Omega J\to 0$. In such a limit, the square of the 
Killing vector $\bchi\equiv\bxi+\Omega\bfeta$ tends to zero on the surface of the body: $\chi^i\chi_i\to 0$. From the 
exterior perspective, this corresponds to the defining condition of a stationary and axisymmetric black hole. 
On its horizon $\cal H$, the Killing vector $\bchi\equiv\bxi+\Omega\bfeta$ is null, with $\Omega$ being the angular 
velocity of the horizon. In the uncharged disc case ($\epsilon=0$), this can immediately be seen from (\ref{Dconstant}) 
and (\ref{M_formula}). Note that $f' \equiv -\chi^i\chi_i$. With charge ($\epsilon\neq 0$) a black hole limit of our
rotating disc occurs if and only if, for $\gamma\to1$, the electric potential in the co--rotating frame $\alpha' \equiv -\chi^i A_i$ tends to a constant on the disc:  
\begin{eqnarray} \label{alpha'_cond}
\gamma\to1: \quad \alpha'(\varrho^*\leq1,\zeta^*=0) = {\rm constant},
\end{eqnarray}
since $\gamma=1$ and $\alpha'= {\rm constant}$ together with (\ref{Dconstant}) 
mean $f'={\rm constant}=f'_{\rm c}=f_{\rm c}=0$ on the disc. Because of the Kerr--Newman black hole uniqueness, see
\cite{Meinel2012}, and the relation
\begin{eqnarray}
 M=2\Omega J+\alpha' Q
\end{eqnarray}
following from (\ref{M_formula}), we are inevitably led to an extreme Kerr--Newman black hole, cf.~\cite{Smarr}. There is 
also another simple argument showing that the black hole limit of the disc must lead to an extreme black hole: 
For the Kerr--Newman black hole, the horizon $\cal H$ is located at a finite interval on the 
$\zeta^\circ$--axis $(\varrho^\circ=0, |\zeta^\circ|\leq l)$, that turns to zero in the extreme case $(l=0)$. This is the 
only possible ``compromise'' with the shape of the disc, and it means in addition, that the coordinate radius $\varrho_0$
of the disc must tend to zero in the limit:
\begin{eqnarray}
\gamma\to1: \quad \varrho_0\to0.
\end{eqnarray}
\par
It should be noted that we are dealing here with the exterior perspective, where the disc shrinks to the origin of the Weyl coordinate system, which also gives the location of the Kerr--Newman black hole's horizon. As discussed by Bardeen and Wagoner (for the uncharged case), the disc region can be considered as a ``singularity'' in the horizon, whereas from the interior perspective (corresponding to finite coordinates $\varrho^*$, $\zeta^*$), the disc remains regular for $\gamma\to 1$ and is surrounded ``by its own infinite, nonasymptotically flat universe'' \cite{BardWag}. For a detailed discusion of these two perspectives in the ECD case, see \cite{Meinel2011}.
\par
In coordinates normalized by the gravitational mass $M$ the extreme Kerr--Newman black hole can be expressed with the parameter $\psi$. The Ernst potentials on the  axis of symmetry are given as
\begin{eqnarray} 
 {\cal E}_{\rm eKN} = \frac{\zeta^\circ-1-\rmi \sqrt{1-\psi^2}}{\zeta^\circ+1-\rmi \sqrt{1-\psi^2}}, \\ 
 {\Phi}_{\rm eKN}  = \frac{\psi}{\zeta^\circ+1-\rmi \sqrt{1-\psi^2}}.
\end{eqnarray}
For that reason we will investigate the post--Newtonian expansion in $\psi$. Interestingly the convergence for the first nine coefficients at $\gamma=1$ is faster in $\psi$ then in $\epsilon$. The extreme Kerr--Newman black hole obeys the parameter relations
\begin{eqnarray}
 & \frac{M\Omega\left(2-\psi^2\right)}{\sqrt{1-\psi^2}}&=1, \label{eKN_rel_1} \\
 & \frac{J^2}{\left(1-\psi^2\right)M^4}&=1, \label{eKN_rel_2}
\end{eqnarray}
and 
\begin{eqnarray} \label{alpha_H}
 \alpha'({\cal H}) =& \frac{\psi}{2-\psi^2}. \label{eKN_rel_3}
\end{eqnarray}

If there is a transition to the extreme Kerr--Newman black hole, then certain functions $R(\gamma,\psi)$ must go to zero for $\gamma\to1$ and $\psi\in[0,1]$. In that case with
\begin{eqnarray}
 \lim\limits_{\gamma\to1}\upsilon(\gamma)=0,
\end{eqnarray}
the inequality
\begin{eqnarray} \label{ineqS}
 \left|\lim\limits_{\gamma\to1}S(\gamma,\psi)\right| = \left|\lim\limits_{\gamma\to1}\frac{R(\gamma,\psi)}{\upsilon(\gamma)}\right|<\infty, \quad \psi\in[0,1]
\end{eqnarray}
or
\begin{eqnarray} \label{ineqS_inv}
 \left|\lim\limits_{\gamma\to1}S^{-1}(\gamma,\psi)\right| = \left|\lim\limits_{\gamma\to1}\frac{\upsilon(\gamma)}{R(\gamma,\psi)}\right|>0, \quad \psi\in[0,1]
\end{eqnarray}
hold for a particular $\upsilon(\gamma)$. It remains to show that $S(\gamma,\psi)$ or $S^{-1}(\gamma,\psi)$ converge at $\gamma=1$. For that we can use the convergence criterion (\ref{suff_cond}) for the majority of cases.
\par

In the majority of cases, we will present the results in a split figure. In part (a) we present the value of a quantity for $\gamma=1$ using increasing expansion orders $N$ to show the convergence behaviour and sometimes using a P\'ade approximation in $\sqrt{\gamma}$ assuming a better convergence then the power series expansion. In part (b) we present the values of the convergence criterion (\ref{suff_cond}) for increasing orders $n$. The critical value $1$ is marked by a dotted line (red). 

\paragraph{Functions on the disc:}
\begin{figure}
  \centering
  \begin{minipage}{0.49\linewidth}
  \subfloat[]{\includegraphics[width=\textwidth]{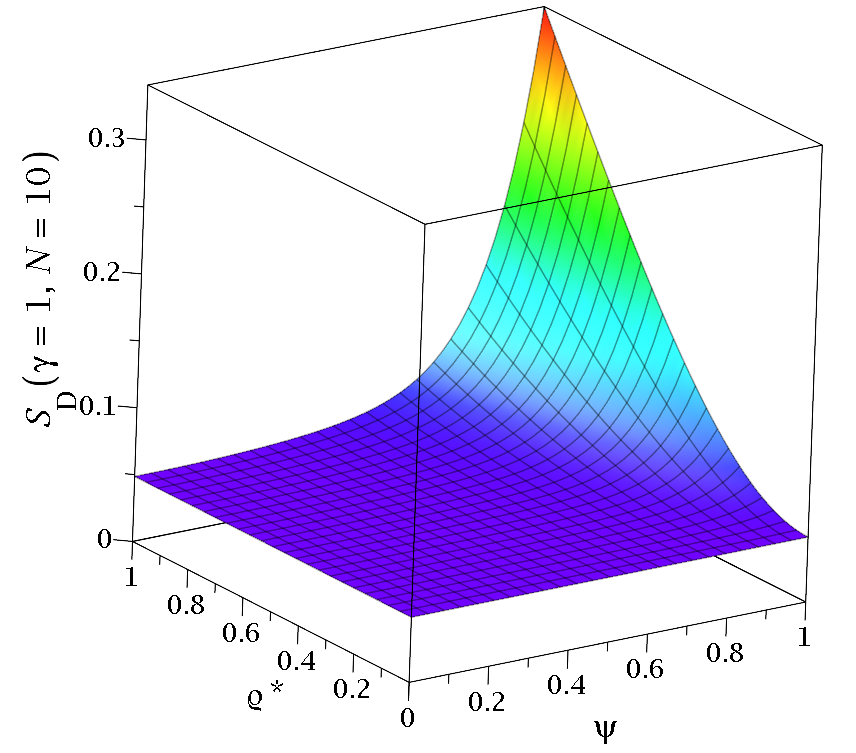}}
  \end{minipage}
  \begin{minipage}{0.49\linewidth}
  \subfloat[]{\includegraphics[width=\textwidth]{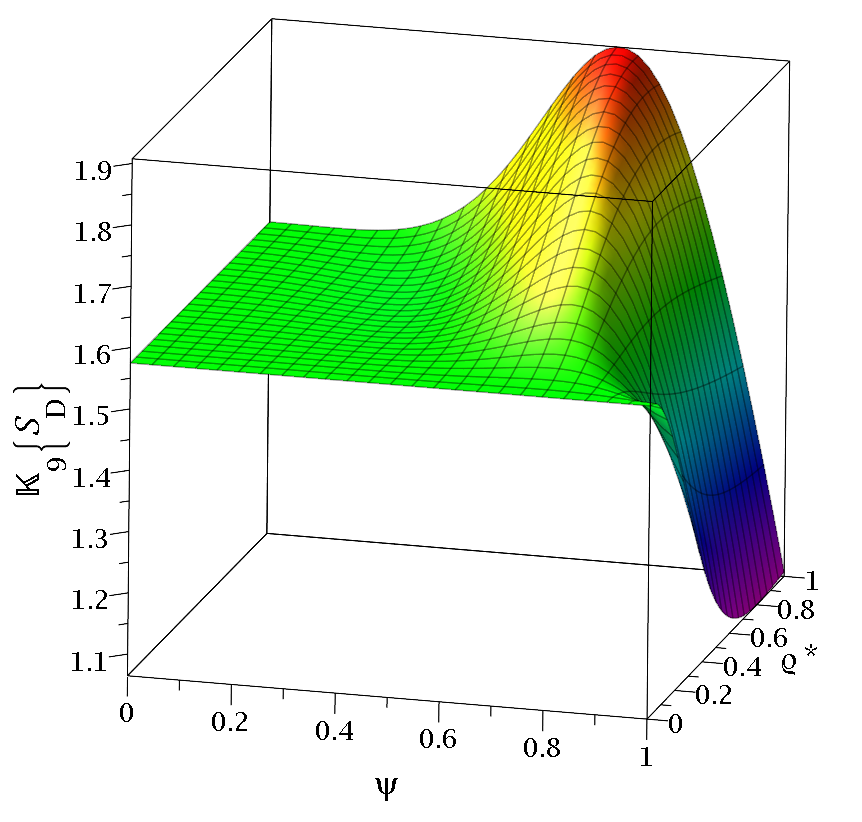}}
  \end{minipage}
  \caption{The quantity $\left[f'(\varrho^*\leq1,\zeta^*=0)\right]^{1/2}\left(1-\gamma\right)^{-1/4}$ in the limit $\gamma\to1$ for the expansion orders $N=10$ (a) and the convergence criterion for $n=9$ (b).}
  \label{SD}
\end{figure}

First of all we want to show that the condition (\ref{alpha'_cond}) is fulfilled. This follows if $f'(\varrho^*\leq1,\zeta^*=0)=0$. For that purpose we investigate
\begin{eqnarray}
\fl S_{\rm D} &= \left[f'(\varrho^*\leq1,\zeta^*=0)\right]^{1/2}\left(1-\gamma\right)^{-1/4} = 1-\left(\frac{3}{4}-\frac{1}{2}\varrho^{*2}\psi^2\right)\gamma + {\cal O}\left(\gamma^2\right) 
\end{eqnarray}
In Figure \ref{SD} (a) we see that $S_{\rm D}$ stays finite on the disc for $\gamma\to1$. Figure \ref{SD} (b) shows that the convergence criterion (\ref{suff_cond}) is fulfilled. For the sake of clarity, we plotted only $S_{\rm D}(\gamma{=}1,N{=}10)$ and $\mathbb{K}_9\{S_{\rm D}\}$. 
\par
\begin{figure}
  \centering
  \begin{minipage}{0.49\linewidth}
  \subfloat[]{\includegraphics[width=\textwidth]{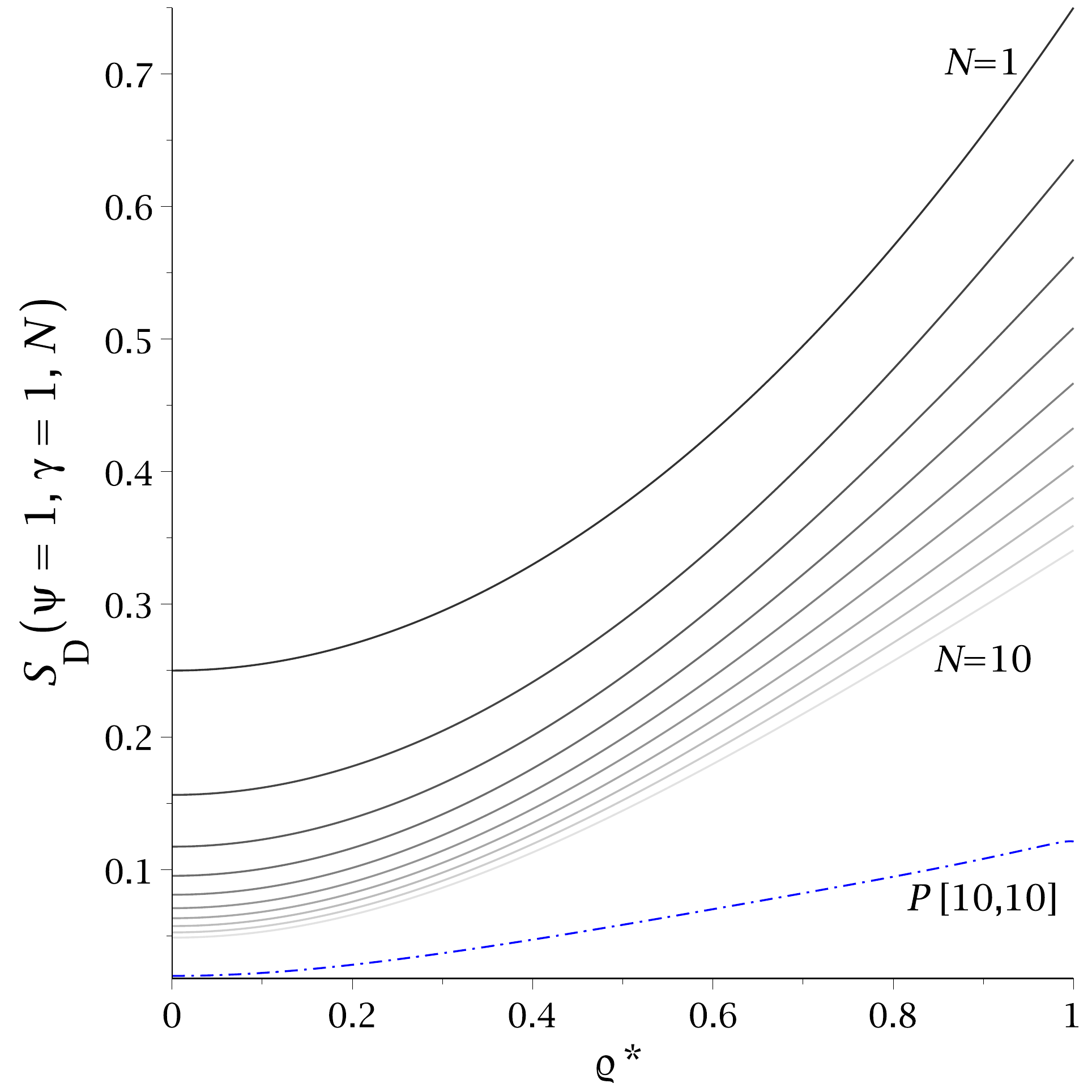}}
  \end{minipage}
  \begin{minipage}{0.49\linewidth}
  \subfloat[]{\includegraphics[width=\textwidth]{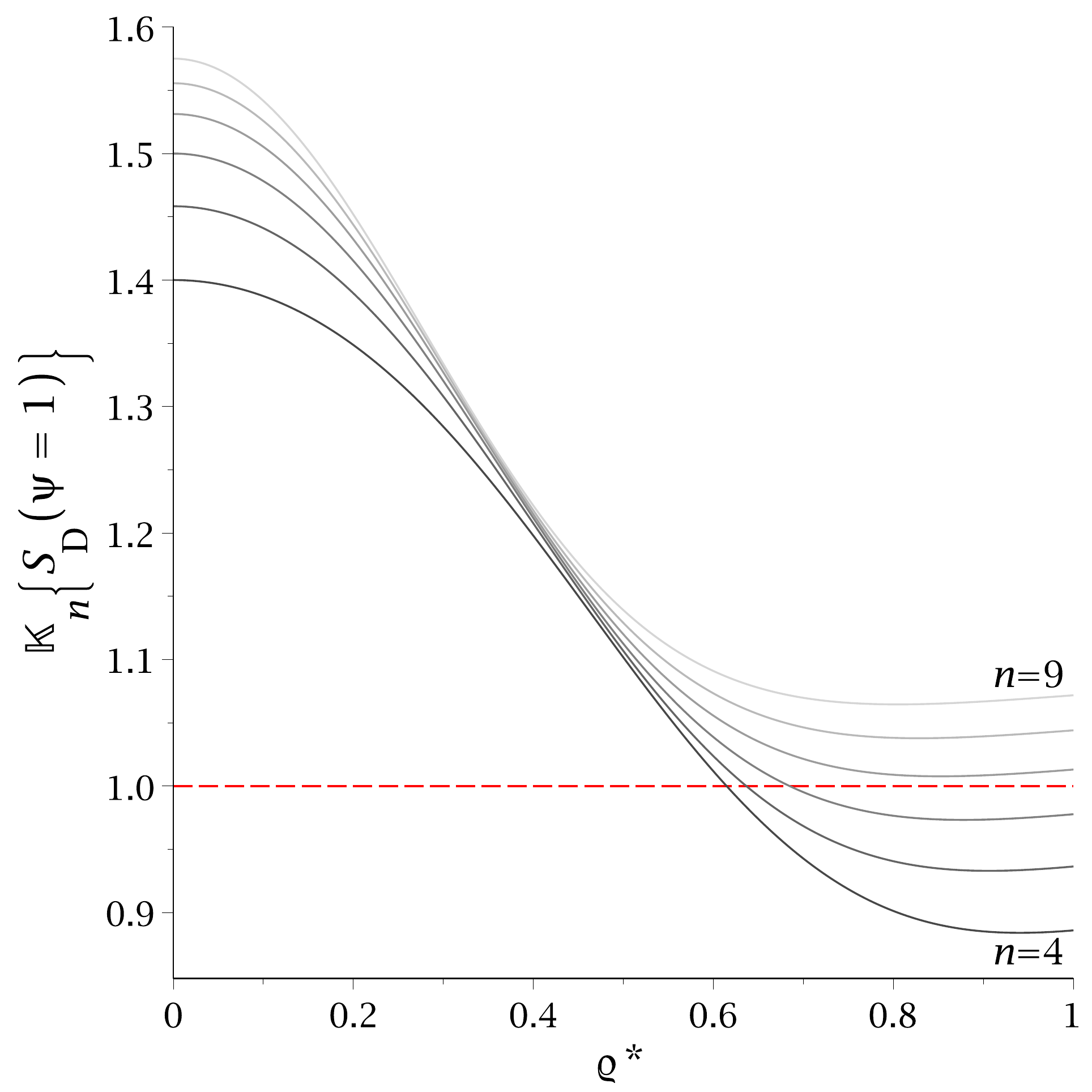}}
  \end{minipage}
  \caption{The quantity $\left[f'(\varrho^*\leq1,\zeta^*=0)\right]^{1/2}\left(1-\gamma\right)^{-1/4}$ for the ECD case $\psi=1$ in the limit $\gamma\to1$ for increasing expansion orders $N$ and for a P\'ade approximation $P[10,10]$ in $\sqrt{\gamma}$ (dashed dotted line)  (a) and the convergence criterion for increasing $n$ (b).}
  \label{SD_ECD}
\end{figure}
The minimum of $\mathbb{K}_9\{S_{\rm D}\}$ is at $\psi=1, \varrho^*\simeq0.8$. The convergence criterion (\ref{suff_cond}) is also fulfilled by $\mathbb{K}_7\{S_{\rm D}\}$ and $\mathbb{K}_8\{S_{\rm D}\}$. The shapes of the plots of $\mathbb{K}_5\{S_{\rm D}\}$ up to $\mathbb{K}_8\{S_{\rm D}\}$ are similar to $\mathbb{K}_9\{S_{\rm D}\}$ with lower values. So $\mathbb{K}_n\{S_{\rm D}\}$ is increasing with higher values of $n$. We plotted the ECD case ($\psi=1$) in Figure \ref{SD_ECD}, were the convergence is at worst. The values for $S_{\rm D}(\psi=1,\gamma=1,N)$ are decreasing with increasing orders $N$. The P\'ade approximation gives a much better result. The convergence criterion is fulfilled. Note that in the ECD case it is sufficient for a black hole limit that $\varrho_0\rightarrow0$ for $\gamma\rightarrow1$.
\par
We have the analytic value $S_{\rm D}(\varrho^*{=}0)=S_{\rm D}(\psi{=}0)=\left(1-\gamma\right)^{3/4}$, and this coincides with Figure \ref{SD} (a) and the values $S_{\rm D}(\varrho^*{=}0,\gamma{=}1,N{=}10)=S_{\rm D}(\psi{=}0,\gamma{=}1,N{=}10)=0.048568...\,$.
\par
Altogether, we find $f'(\varrho^*\leq1,\zeta^*=0)=0$ and with (\ref{Dconstant}) that (\ref{alpha'_cond}) is fulfilled.

\paragraph{Disc radius:}

\begin{figure}
  \centering
  \begin{minipage}{0.49\linewidth}
  \subfloat[]{\includegraphics[width=\textwidth]{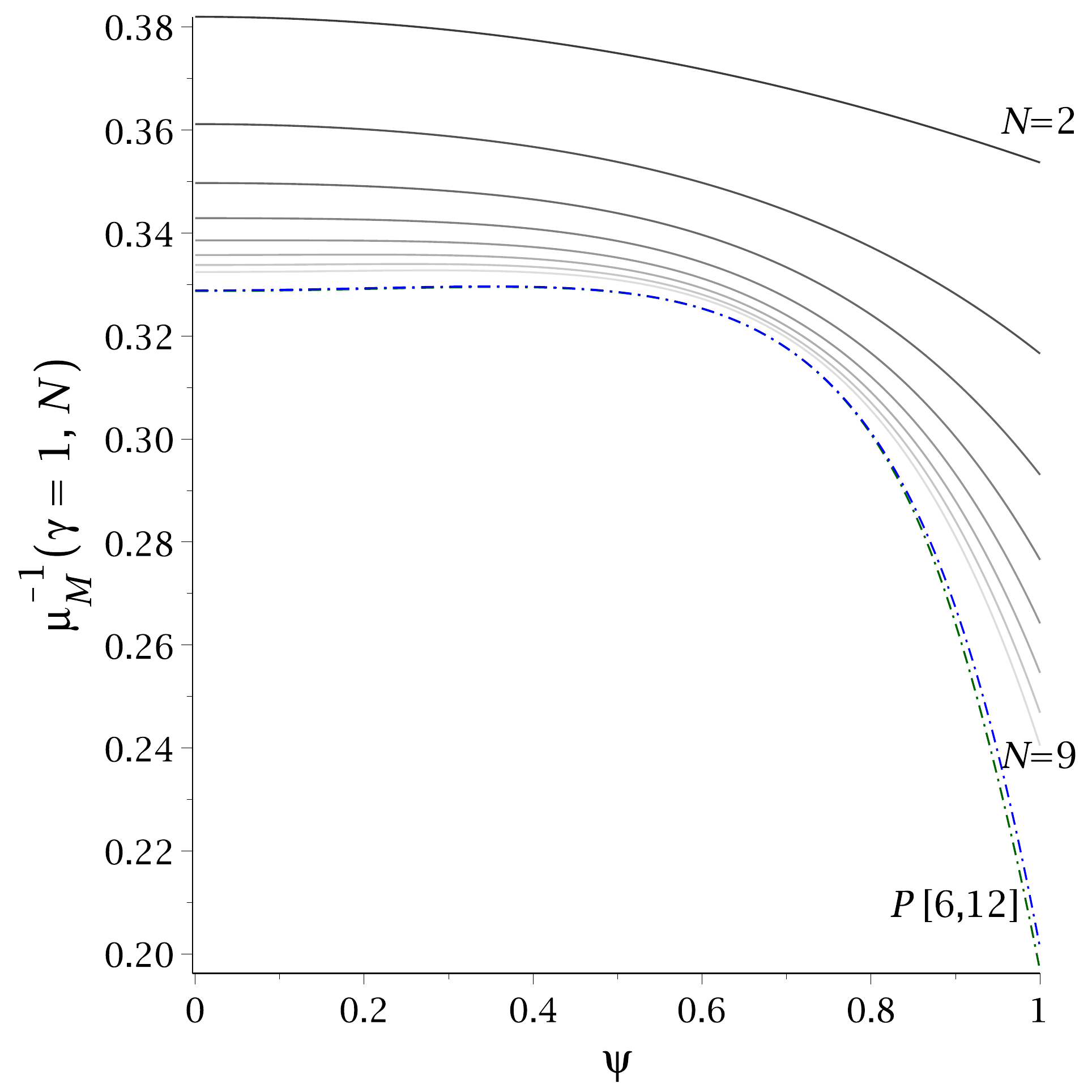}}
  \end{minipage}
  \begin{minipage}{0.49\linewidth}
  \subfloat[]{\includegraphics[width=\textwidth]{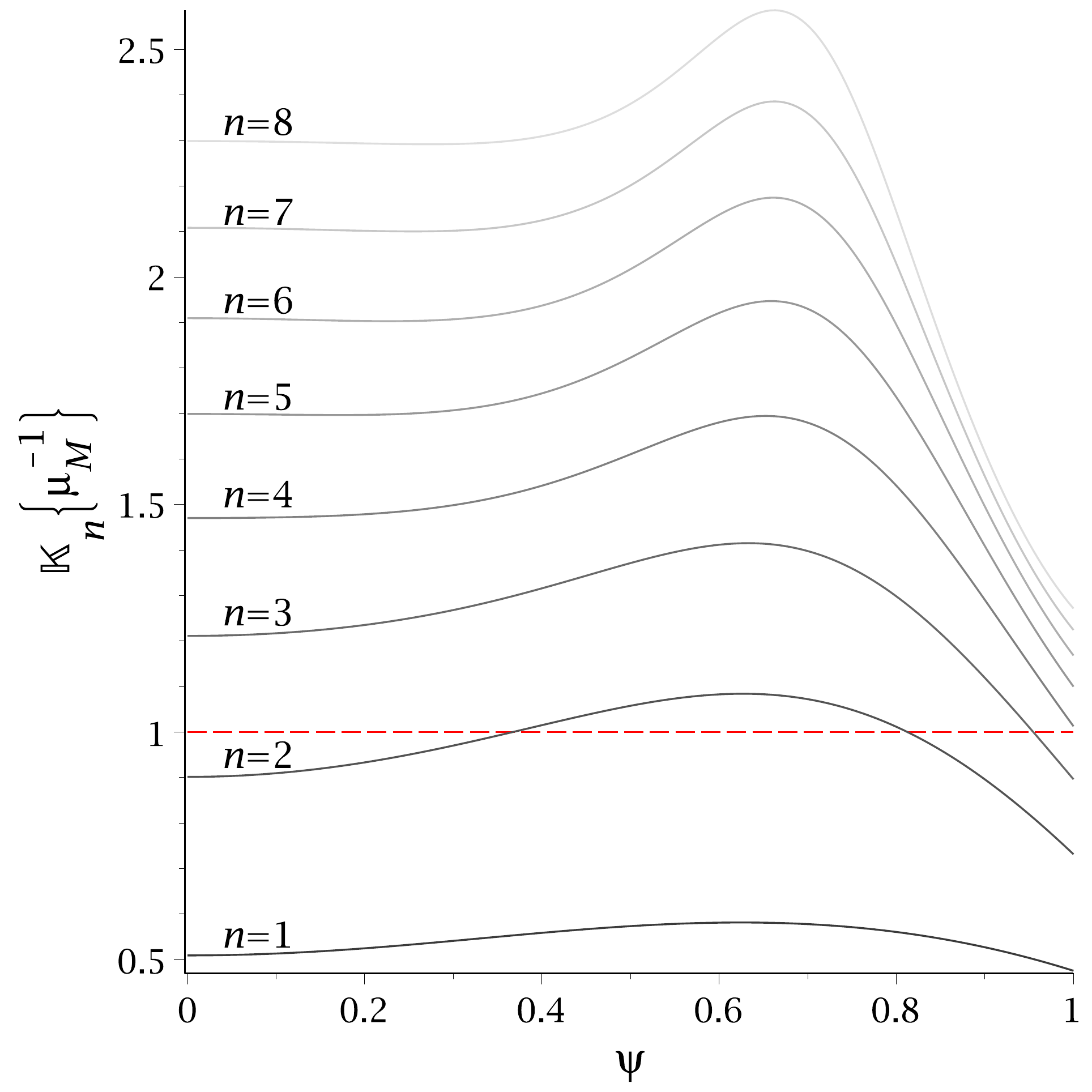}}
  \end{minipage}
  \caption{The ratio $M^*(1-\gamma)$ in the limit $\gamma\to1$ for increasing expansion orders $N$ and for two P\'ade approximations $P[6,12]$ in $\sqrt{\gamma}$ (dashed dotted lines, blue/green) (a) and the convergence criterion for increasing $n$ (b).}
  \label{mu_inv_M}
\end{figure}

In order to show, that the disc radius $\varrho_0$ vanishes for $\gamma\to1$, we will investigate the ratio
\begin{eqnarray}
 \mu_M^{-1} = \frac{M}{\varrho_0}(1-\gamma) = \frac{4}{3\pi}\gamma\left[1+\left(\frac{9}{10}-\frac{1}{15}\psi^2\right)\gamma + {\cal O}\left(\gamma^2\right)\right].
\end{eqnarray}
In Figure \ref{mu_inv_M} (a) we can see that the inequality
\begin{eqnarray}
 \lim\limits_{\gamma\to1}\mu_M^{-1}>0,\quad \psi\in[0,1]
\end{eqnarray}
of type (\ref{ineqS_inv}) is fulfilled. Figure \ref{mu_inv_M} (b) shows, that the $\mathbb{K}_n$ from (\ref{suff_cond}) exceed one for large values of $n$. 
\par
The analytic value in the uncharged case is 
\[\mu_{M}^{-1}(\gamma=1,\psi=0)=0.32863...\,,\]
while the Pad\'e approximation (blue dashed dotted line in figure \ref{mu_inv_M} (a)) gives 
\[P[6,12]\{\mu_{M}^{-1}(\gamma=1,\psi=0,N{=}9)\}=0.32885...\,.\]
\par
We also expanded $\mu_M$ in $\gamma$ and plotted the inverse P\'ade approximation $\left(P[6,12]\{\mu_{M}(\gamma=1)\}\right)^{-1}$ as a green dashed dotted line in figure \ref{mu_inv_M} (a). To compare the values, we have:
\begin{eqnarray}
\left(P[6,12]\{\mu_{M}(\gamma=1,\psi=0,N{=}9)\}\right)^{-1}=0.32872...\,, \nonumber\\
\left(P[6,12]\{\mu_{M}(\gamma=1,\psi=1,N{=}9)\}\right)^{-1}=0.19675...\,, \nonumber \\
P[6,12]\{\mu_{M}^{-1}(\gamma=1,\psi=1,N{=}9)\}=0.20117...\,. \nonumber
\end{eqnarray}
\par
All in all we find that in the limit $\gamma\to1$ the disc radius $\varrho_0$ tends to zero.

\paragraph{Parameter relations from the extreme Kerr--Newman black hole:}
\begin{figure}
  \centering
  \begin{minipage}{0.49\linewidth}
  \subfloat[]{\includegraphics[width=\textwidth]{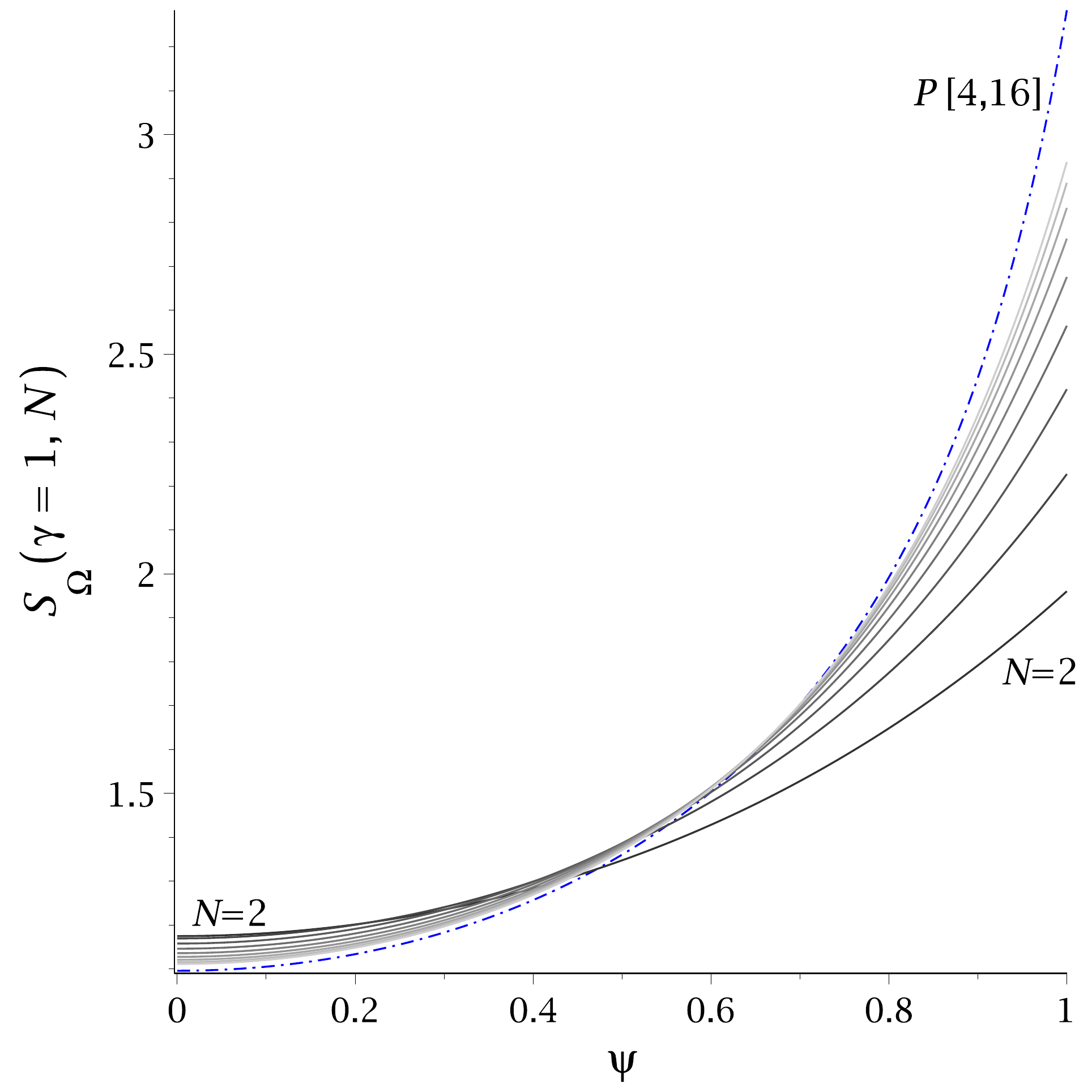}}
  \end{minipage}
  \begin{minipage}{0.49\linewidth}
  \subfloat[]{\includegraphics[width=\textwidth]{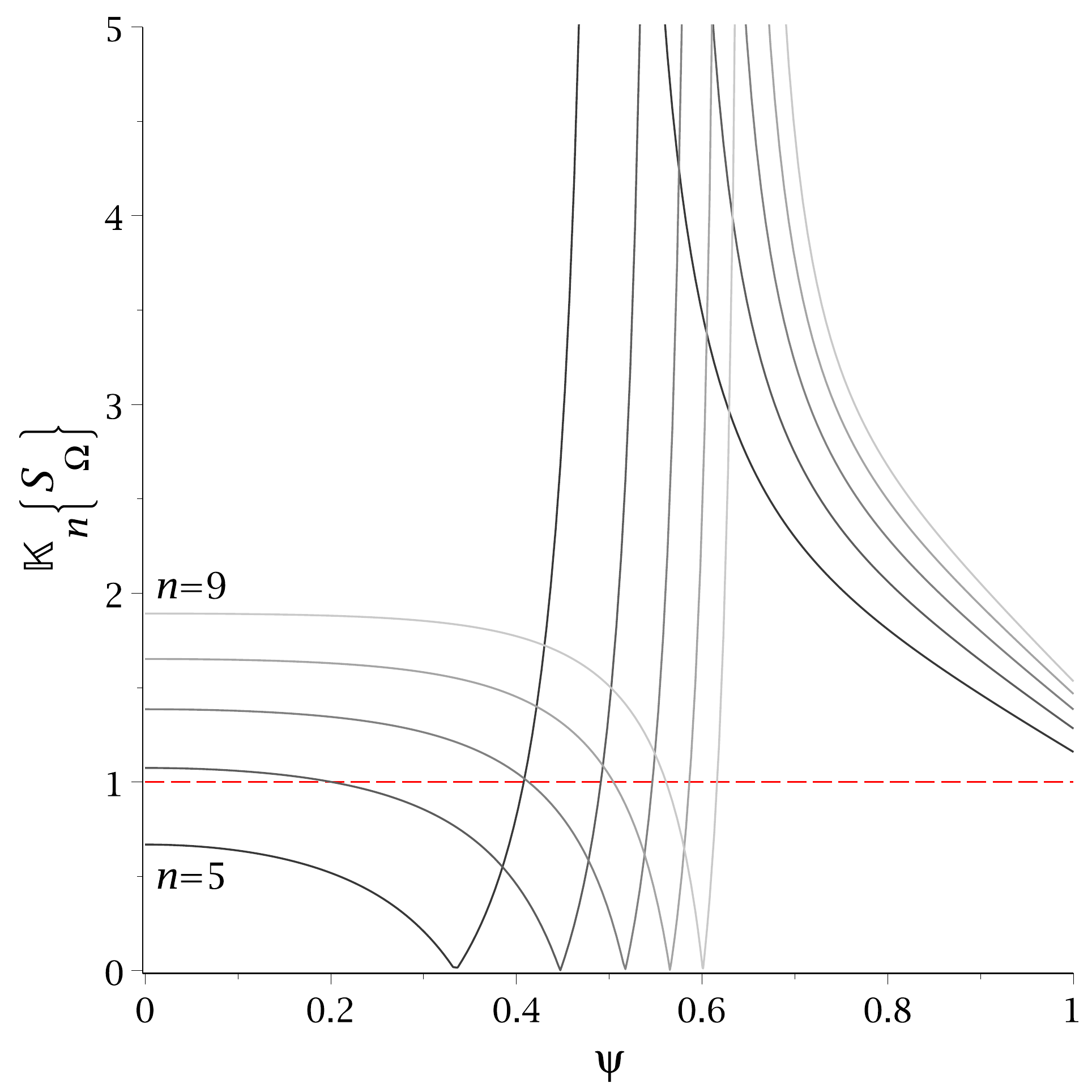}}
  \end{minipage}
  \begin{minipage}{0.49\linewidth}
  \subfloat[]{\includegraphics[width=\textwidth]{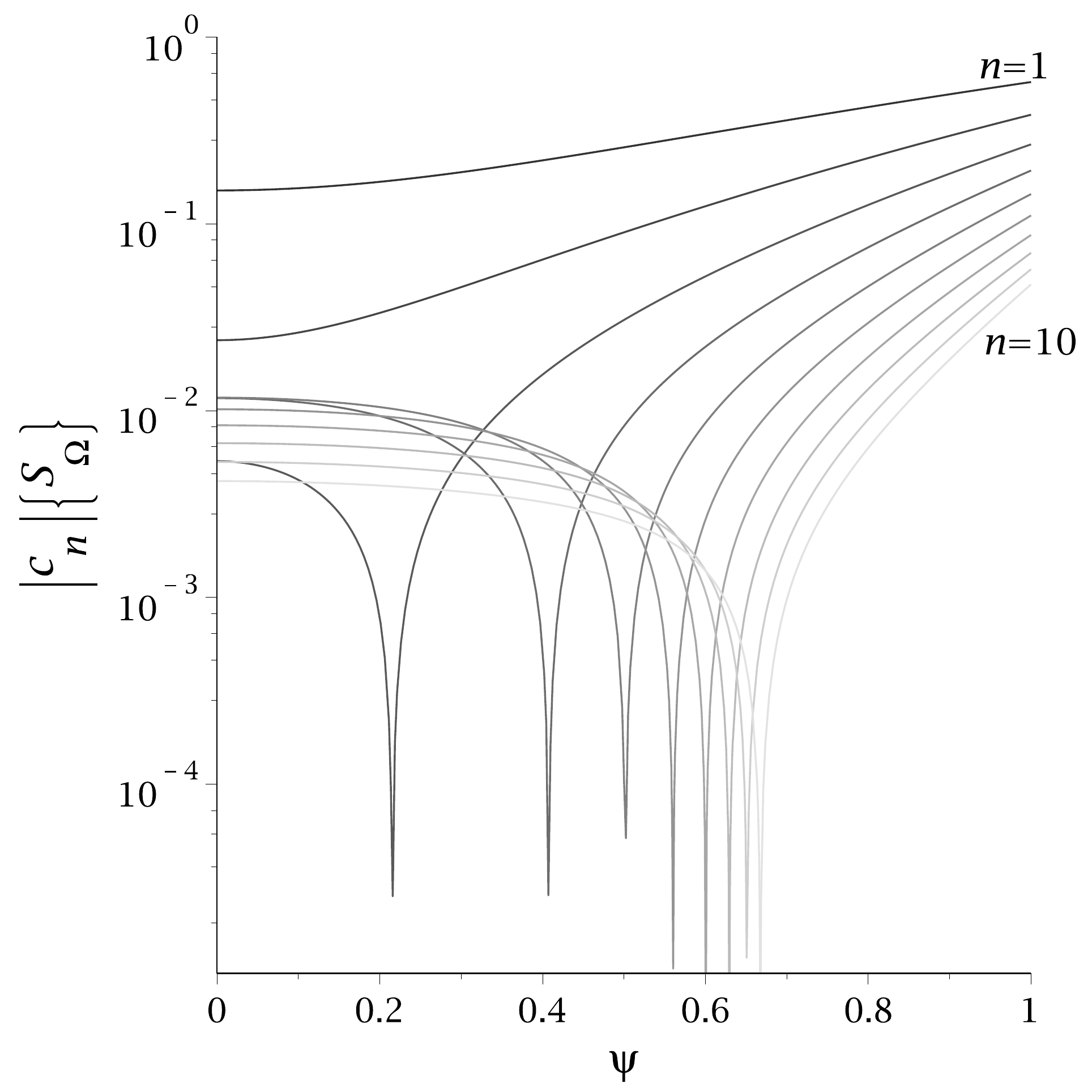}}
  \end{minipage}
  \caption{The quantity $S_\Omega$ in the limit $\gamma\to1$ for increasing expansion orders $N$ and for a P\'ade approximation $P[4,16]$ in $\sqrt{\gamma}$ (dashed dotted line) (a), the convergence criterion for increasing $n$ (b) and the coefficient functions $|c_n|$ (c).}
  \label{SOm}
\end{figure}
Now we check the parameter relations (\ref{eKN_rel_1}) and (\ref{eKN_rel_2}). Note that $\Omega$ and $J$ have the global pre--factor $\sqrt{1-\psi^2}$. At first we will investigate the quantity
\begin{eqnarray}
\fl S_\Omega = \left[1-\frac{M\Omega\left(2-\psi^2\right)}{\sqrt{\gamma(1-\psi^2)}}\right]\left(1-\gamma\right)^{-1} = 1+\left(1-\frac{8}{3 \pi} + \frac{4}{3\pi}\,\psi^2\right)\gamma + {\cal O}\left(\gamma^2\right).
\end{eqnarray}
In Figure \ref{SOm} (a) we can see that $S_\Omega$ stays finite at $\gamma=1$. The singular behaviour of the $\mathbb{K}_n$ around $\psi=0.5$ (b) can be explained by the zeros of the coefficient functions $c_n$ (c). Although (\ref{suff_cond}) is not valid in a small interval in $\psi$, the convergence of $S_\Omega$ is extremely good there, which can be seen in (a). 
\par
\begin{figure}
  \centering
  \begin{minipage}{0.49\linewidth}
  \subfloat[]{\includegraphics[width=\textwidth]{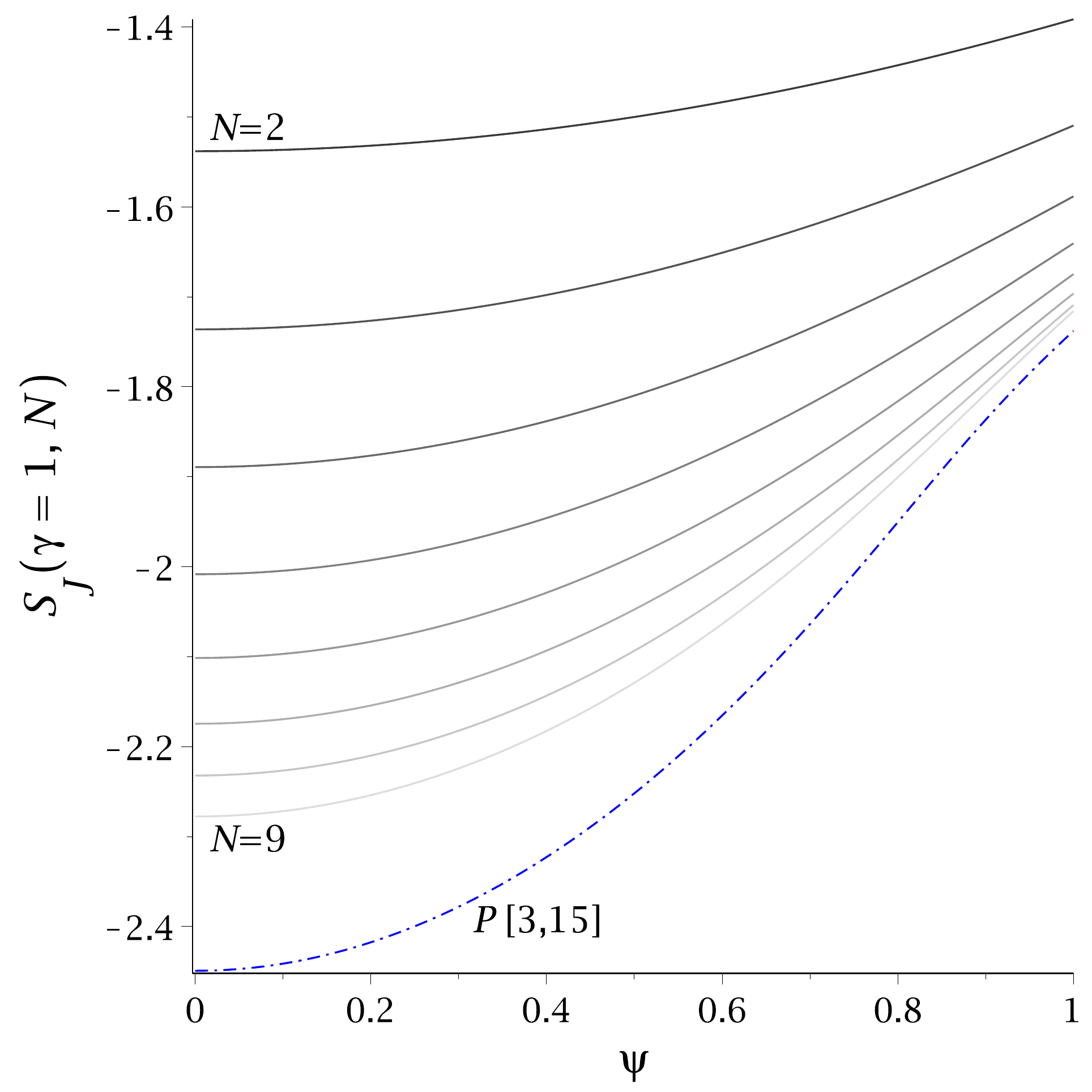}}
  \end{minipage}
  \begin{minipage}{0.49\linewidth}
  \subfloat[]{\includegraphics[width=\textwidth]{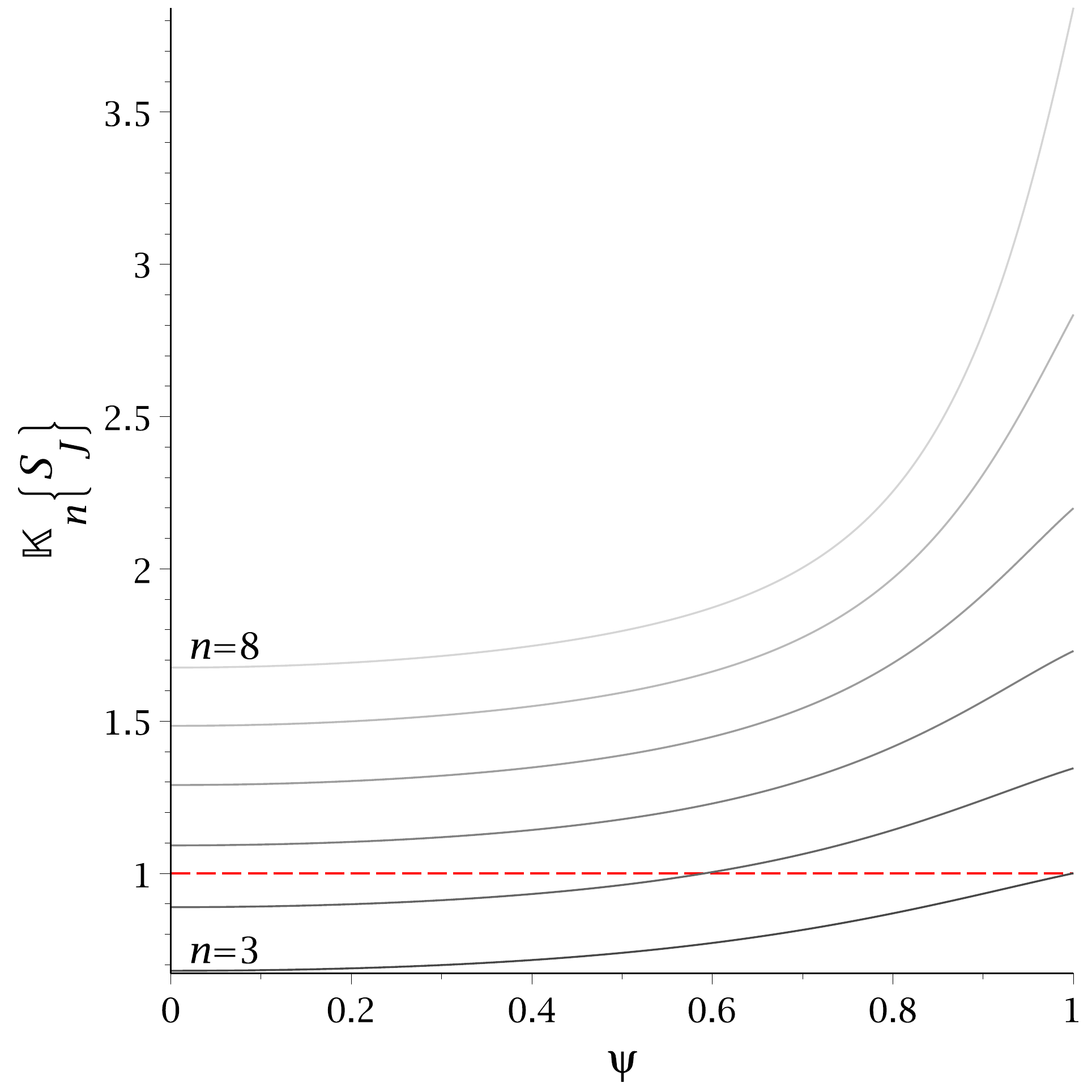}}
  \end{minipage}
  \caption{The quantity $\left[1/2+\left[1-J^{\circ}/\sqrt{\gamma\left(1-\psi^2\right)}\right]\left(1-\gamma\right)^{-1}\right]\left(1-\gamma\right)^{-1}$ in the limit $\gamma\to1$ for increasing expansion orders $N$ and for a P\'ade approximation $P[3,15]$ in $\sqrt{\gamma}$ (dashed dotted line) (a) and the convergence criterion for increasing $n$ (b).}
  \label{SJ}
\end{figure}
It turns out that the quantity
\begin{eqnarray}
S_J &= \left[ \frac{1}{2} + \left(1-\frac{J^\circ}{\sqrt{\gamma\left(1-\psi^2\right)}}\right)\left(1-\gamma\right)^{-1} \right]\left(1-\gamma\right)^{-1} \\ &= -\frac{3\pi}{10\gamma}\left[1+\left(\frac{39}{20} - \frac{5}{\pi} - \frac{1}{14}\,\psi^2\right)\gamma + {\cal O}\left(\gamma^2\right)  \right]
\end{eqnarray}
converges and stays finite in the limit $\gamma\to1$ (see Figure \ref{SJ}). This is a check for (\ref{eKN_rel_2}) and moreover it means that $J^\circ$ can be written as
\begin{eqnarray} \label{Lob_J}
J^\circ = \sqrt{1-\psi^2}\left[1 + \left(c_2-3/8\right)\left(1-\gamma\right)^2 + {\cal O}\left(1-\gamma\right)^3\right],
\end{eqnarray}
with $\lim\limits_{\gamma\to1}S_J=-c_2$. We will use this result in section \ref{sec:LOB_BL}.
\par
If (\ref{eKN_rel_1}) and (\ref{eKN_rel_2}) hold for the disc at $\gamma=1$, it follows with (\ref{M_formula}) that
\begin{eqnarray}
 \lim\limits_{\gamma\to1}\alpha'_{\rm c}=\frac{\psi}{2-\psi^2}.
\end{eqnarray}
In accordance with (\ref{alpha_H}) we identify the constant in (\ref{alpha'_cond}) to be
\begin{eqnarray}
\lim\limits_{\gamma\to1}\alpha'(\varrho^*\leq1,\zeta^*=0) = \psi\left(2-\psi^2\right)^{-1}.
\end{eqnarray}

\paragraph{Conclusion:}
Within the high accuracy of these calculations we can state that the solution describing rigidly rotating discs of charged dust has a transition to an extreme Kerr--Newman black hole. The multipole moments also fit in this result, as implied in the next section.

\section{Leading order behaviour close to the black hole limit} \label{sec:LOB_BL}
Interestingly, it turns out that the power series expansion at $\gamma=0$ could be used to make a statement about the leading order behaviour for the power series expansion at $\gamma=1$. The uncharged case is discussed analytically in \cite{KLM}. For this purpose we investigated the coefficients from (\ref{def_mn_en}) in two normalized and dimensionless forms:
\begin{eqnarray}
 m_n^\circ = \frac{m_n}{m_0^{n+1}} \quad{\rm and}\quad e_n^\circ = \frac{e_n}{m_0^{n+1}} \quad{\rm with}\quad m_0 = M, \quad m_1 = \rmi J,
\end{eqnarray}
\begin{eqnarray}
\fl m_n^\star = \frac{m_n}{k_n}\left(1-\psi^2\right)^{n/2} \quad{\rm and}\quad e_n^\star = \frac{e_n}{\psi k_n}\left(1-\psi^2\right)^{n/2} \quad{\rm with}\quad k_n = m_0^{n+1}m_1^{\circ n}.
\end{eqnarray}
Herein the $k_n$ correspond exactly to the coefficients $m_n$ and accordingly $e_n/\psi$ of a Kerr--Newman solution with mass $M$, angular momentum $J$ and charge Q. Thus $m_n^\star = e_n^\star = \left(1-\psi^2\right)^{n/2}$ for black holes, and they are equal to $m_n^\circ$ and accordingly $e_n^\circ/\psi$ in the extreme case.
\par
\begin{figure}
  \centering
  \begin{minipage}{0.49\linewidth}
  \subfloat[]{\includegraphics[width=\textwidth]{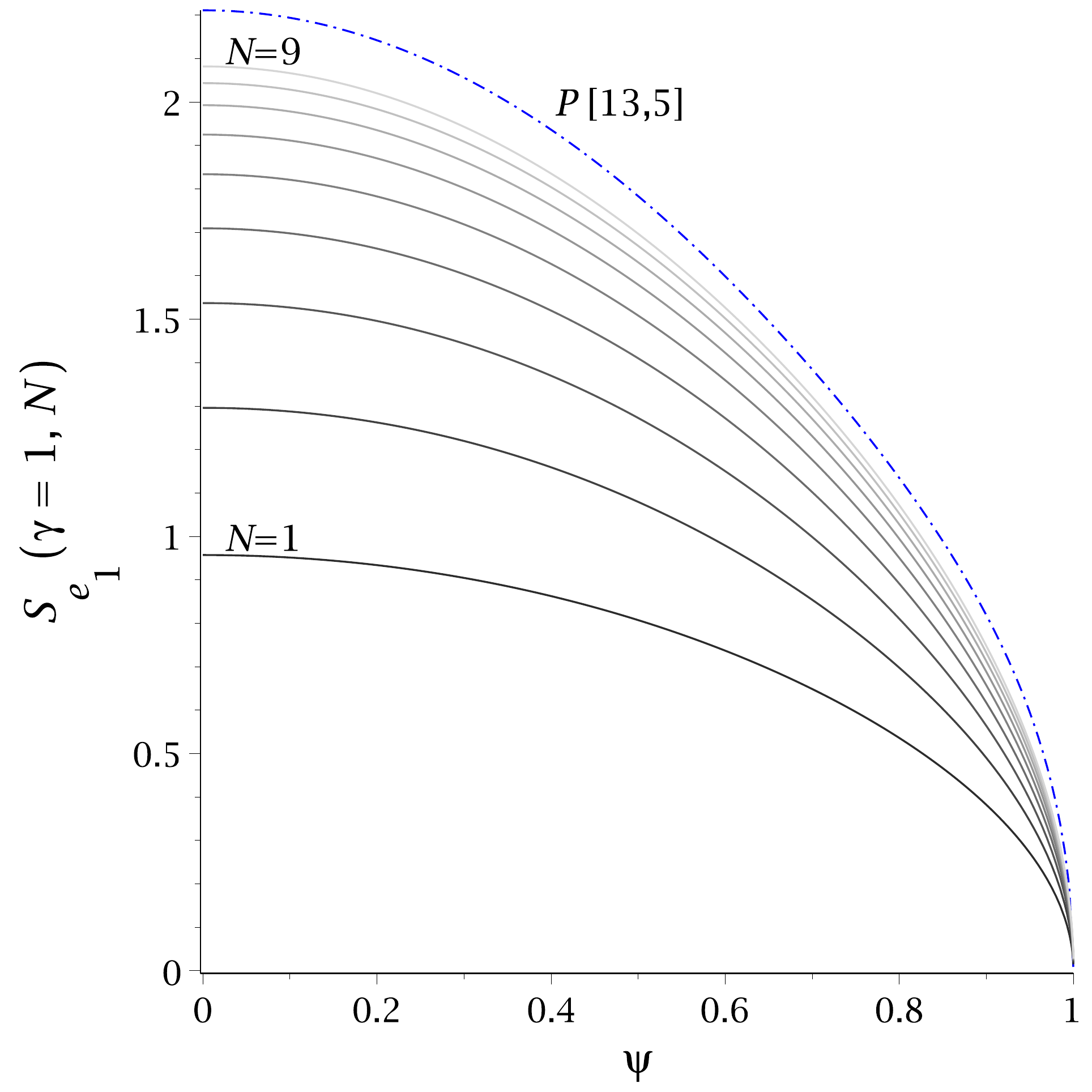}}
  \end{minipage}
  \begin{minipage}{0.49\linewidth}
  \subfloat[]{\includegraphics[width=\textwidth]{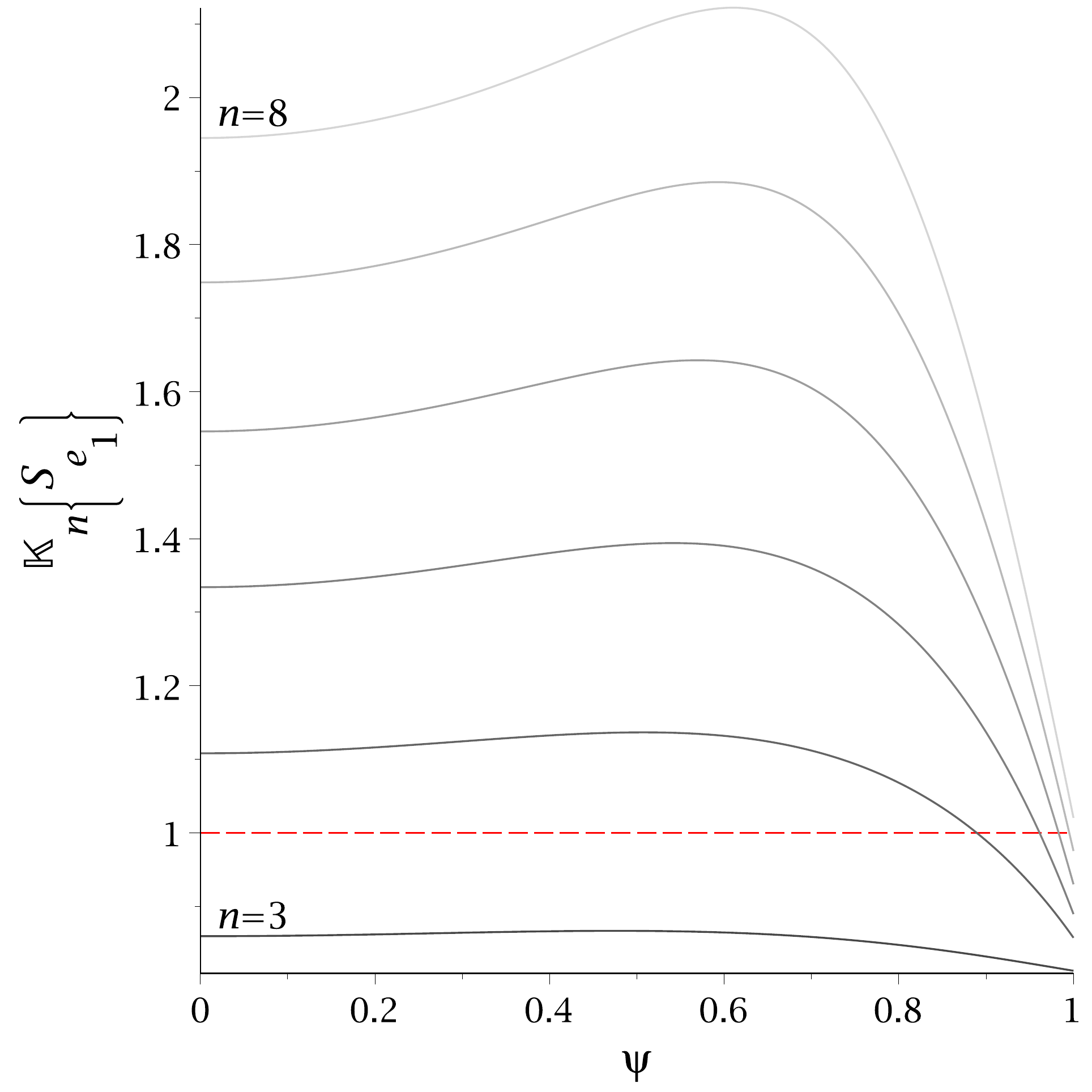}}
  \end{minipage}
  \caption{The quantity $\left[\sqrt{1-\psi^2}-e^\star_1\right]\left(1-\gamma\right)^{-2}$ in the limit $\gamma\to1$ for increasing expansion orders $N$ and for a P\'ade approximation $P[13,5]$ in $\sqrt{\gamma}$ (dashed dotted line) (a) and the convergence criterion for increasing $n$ (b).}
  \label{Se1}
\end{figure}
In the black hole limit we checked
\begin{eqnarray} \label{mn_en_diff}
 \frac{\rmd m_n^\circ}{\rmd \gamma}\Biggl|_{\gamma=1} = \frac{\rmd e_n^\circ}{\rmd \gamma}\Biggl|_{\gamma=1} = 0, \quad \frac{\rmd m_n^\star}{\rmd \gamma}\Biggl|_{\gamma=1} = \frac{\rmd e_n^\star}{\rmd \gamma}\Biggl|_{\gamma=1} = 0
\end{eqnarray}
up to $n=11$, while the second derivatives of the $m_n^\star$ at $\gamma=1$ are only zero in the uncharged case. We already showed that (\ref{mn_en_diff}) holds for $m_1^\circ$. It follows with (\ref{Lob_J}) that
\begin{eqnarray}
 \frac{\rmd m_n^\circ}{\rmd \gamma}\Biggl|_{\gamma=1} = 0 \Leftrightarrow \frac{\rmd m_n^\star}{\rmd \gamma}\Biggl|_{\gamma=1} = 0, \quad \frac{\rmd e_n^\circ}{\rmd \gamma}\Biggl|_{\gamma=1} = 0 \Leftrightarrow \frac{\rmd e_n^\star}{\rmd \gamma}\Biggl|_{\gamma=1} = 0
\end{eqnarray}
for any fixed value of $n$. This is important, because some quantities have a better (faster) convergence then others.
\par
\begin{figure}
  \centering
  \begin{minipage}{0.49\linewidth}
  \subfloat[]{\includegraphics[width=\textwidth]{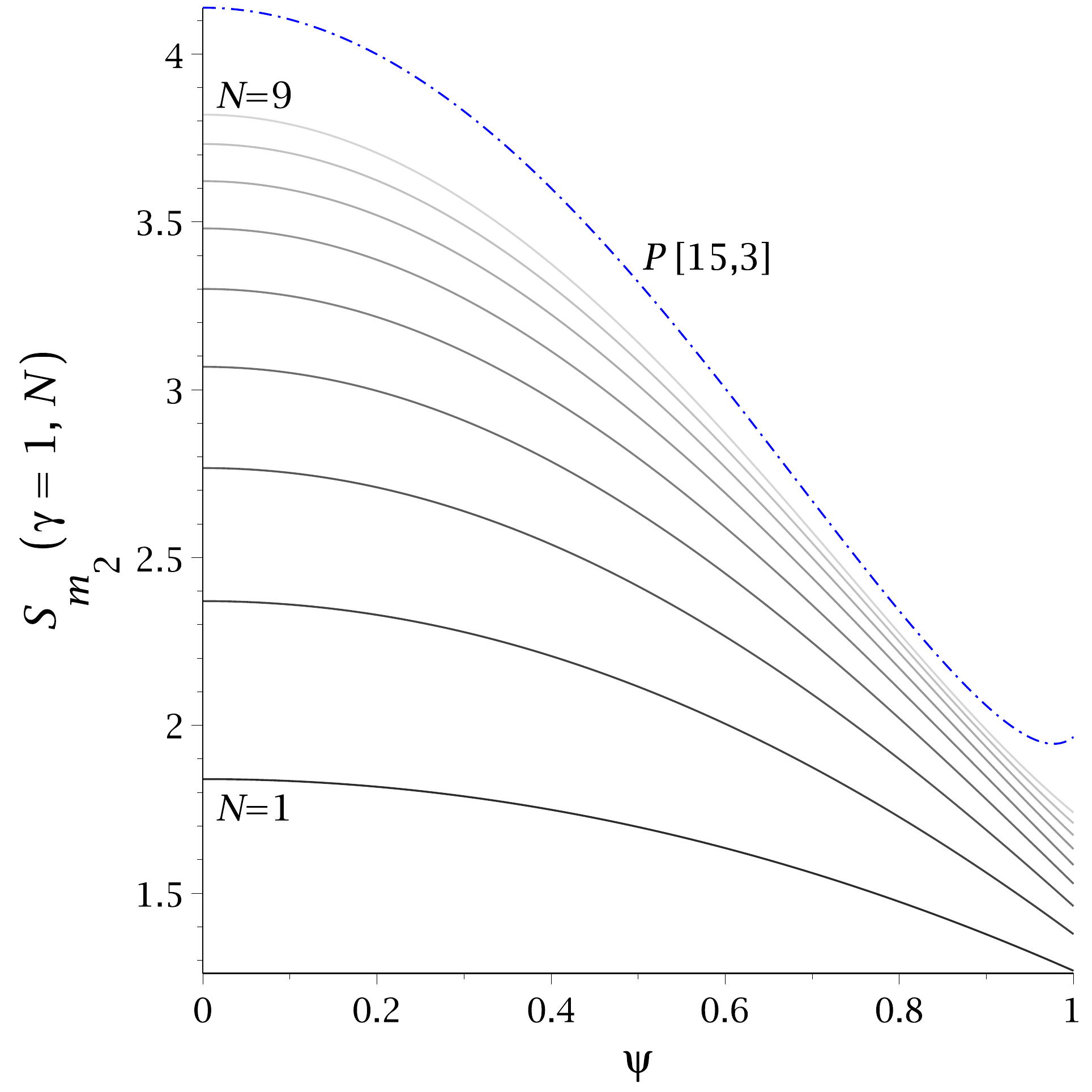}}
  \end{minipage}
  \begin{minipage}{0.49\linewidth}
  \subfloat[]{\includegraphics[width=\textwidth]{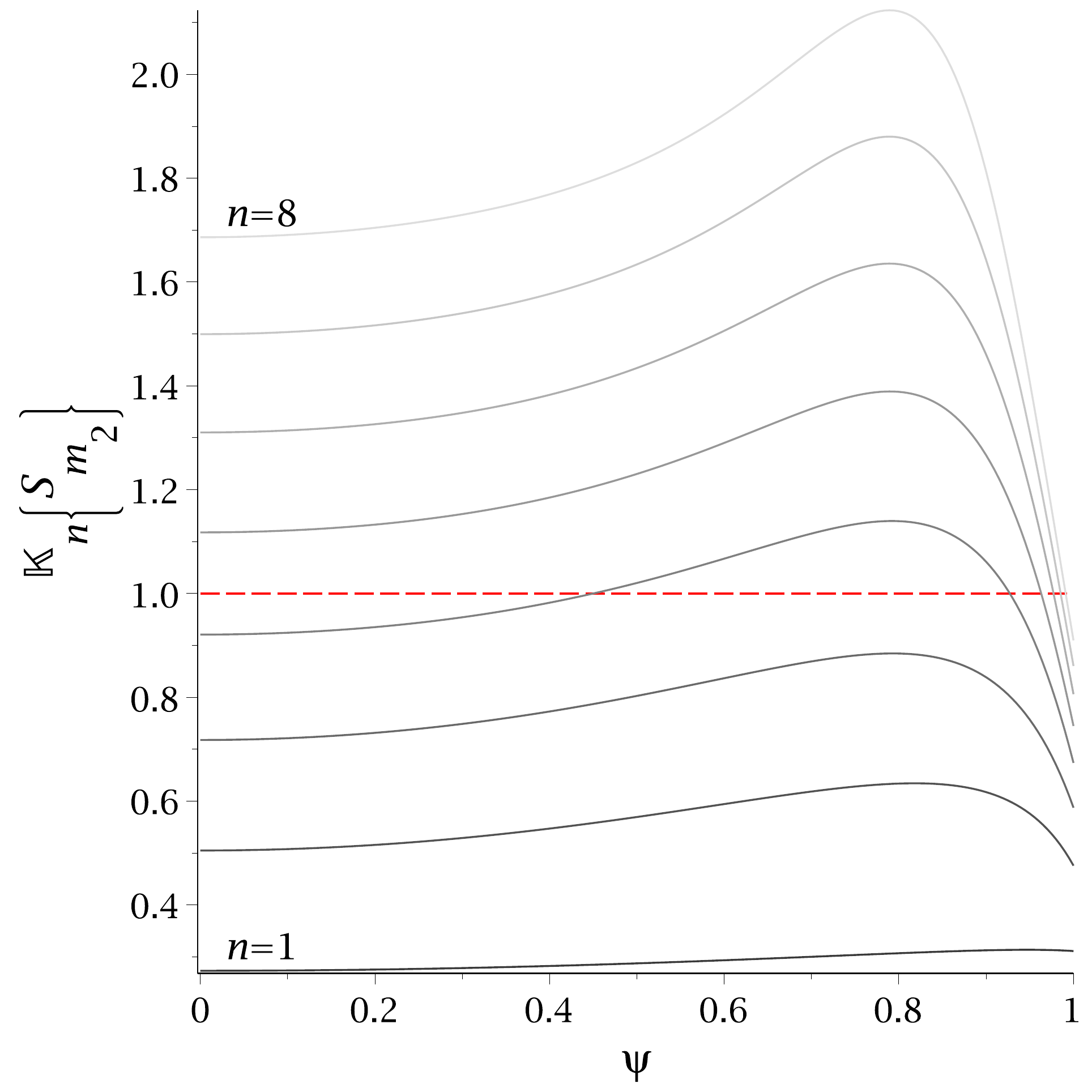}}
  \end{minipage}
  \caption{The quantity $\left[\rmi^2\left(1-\psi^2\right)-m^\circ_1\right]\left(1-\gamma\right)^{-2}$ in the limit $\gamma\to1$ for increasing expansion orders $N$ and for a P\'ade approximation $P[15,3]$ in $\sqrt{\gamma}$ (dashed dotted line) (a) and the convergence criterion for increasing $n$ (b).}
  \label{Sm2}
\end{figure}

We present the following two important quantities as examples: The magnetic dipole moment $e^\star_1$ and the gravitational quadrupole moment $m^\circ_2$, namely we investigate
\begin{eqnarray}
\fl S_{e_1} = \left[\sqrt{1-\psi^2}-e^\star_1\right]\left(1-\gamma\right)^{-2} = \frac{1}{2}\sqrt{1-\psi^2}\left[1+\left(\frac{32}{35}-\frac{1}{5}\,\psi^2\right)\gamma + {\cal O}\left(\gamma^2\right)\right]
\end{eqnarray}
and
\begin{eqnarray}
\fl S_{m_2} = \left[\rmi^2\left(1-\psi^2\right)-m^\circ_1\right]\left(1-\gamma\right)^{-2} = \frac{9\pi^2}{80\gamma^2}\left[1+\left(\frac{23}{35}-\frac{18}{35}\,\psi^2\right)\gamma + {\cal O}\left(\gamma^2\right)\right].
\end{eqnarray}
In Figure \ref{Se1} (a) and \ref{Sm2} (a) we see that $S_{e_1}$ and $S_{m_2}$ stay finite in the limit $\gamma\to1$. The convergence criterion for $S_{e_1}$ is fulfilled in the regular parameter space and for $S_{m_2}$ up to $|\psi|\leq0.99$ at $n=8$ (see Figure \ref{Sm2} (b)). We find a similar convergence behaviour for the electric quadrupole moment $e_2^\circ$, while for higher multipole moments, up to $n=11$, the convergence criterion is fulfilled in the entire regular parameter space.
\par
For the following discussion we assume that (\ref{mn_en_diff}) holds for all $n\geq0$. If this is the case the Ernst potentials ${\cal E}$ and $\Phi$ can be expanded at $\gamma=1$ in the following way: The series starts with the Ernst potentials ${\cal E}_{\rm KN}$ and $\Phi_{\rm KN}$ of the Kerr--Newman solution with mass $M$, charge $Q$ and angular momentum $J$ and the first order term in $\left(1-\gamma\right)$ is zero:
\begin{eqnarray}
 {\cal E}(M,J,Q;\zeta)        = {\cal E}_{\rm KN}(M,J;\zeta) + {\cal O}\left[\left(1-\gamma\right)^2 \right], \\ 
 {\cal E}_{\rm KN}(M,J;\zeta) = \frac{M(\zeta-M)-\rmi J}{M(\zeta+M)-\rmi J}.
\end{eqnarray}
\begin{eqnarray}
 {\Phi}(M,J,Q;\zeta)          = {\Phi}_{\rm KN}(M,J,Q;\zeta) + {\cal O}\left[\left(1-\gamma\right)^2 \right], \\ 
 {\Phi}_{\rm KN}(M,J,Q;\zeta)  = \frac{QM}{M(\zeta+M)-\rmi J}.
\end{eqnarray}
Comparing (\ref{Lob_J}) with the power series expansion from $J^\circ$ at $\gamma=0$, we can state that
\begin{eqnarray}
 \frac{Q^2}{M^2} + \frac{J^2}{M^4} \geq 1, \quad \gamma\in[0,1], \; \epsilon\in[-1,1].
\end{eqnarray}
It is equal to one if $\gamma=1$ or $\psi=\pm1$. This means that ${\cal E}_{\rm KN}$ and $\Phi_{\rm KN}$ are given by the axis potential of a ``hyper\-extreme'' Kerr--Newman solution for $\gamma<1$ and $|\psi|<1$. The Kerr--Newman solution already contains the first multipole moments $M,J$ and $Q$. Therefore, in the far field the second order term in $(1-\gamma)$ for $\cal E$ starts with order $\zeta^{-3}$ whereas the term for $\Phi$ starts with order $\zeta^{-2}$.
\par
All together we found strong evidence that the disc solution near the ultra--relativistic limit at $\gamma=1$, and not to close to the disc itself, could be approximated very well by a ``hyper\-extreme'' Kerr--Newman solution with the same gravitational mass, angular momentum and charge.
 
\ack{This research was supported by the Deutsche Forschungsgemeinschaft (DFG) through the Graduiertenkolleg 1523 ``Quantum and Gravitational Fields''.}

\appendix
\section*{Appendix: Coefficient functions} \label{APP}
General structure of an approximated function (see (\ref{gen_pow_ser_approx})):
\begin{eqnarray}
 F(\gamma,\psi,N) = F_{\rm NL}(\gamma,\psi)\left[1 + \sum\limits_{n=1}^N c_n(\psi)\gamma^n\right].
\end{eqnarray}

\begin{itemize}
 \item Charge parameter: $\psi = Q/M = \epsilon\left(1-E_{\rm B}^{(\rm rel)}\right)^{-1}$. For the relative binding energy $E_{\rm B}^{(\rm rel)}$ see Appendix C. in \cite{MeinPal}.
 \item Funtion on the disc $f'(\varrho^*\leq1,\zeta^*=0)$:
       \begin{eqnarray}
        \fl f'_{\rm NL} = 1, \\
        \fl c_1 = -2 + {\psi}^{2}{\varrho}^{2}, \\
        \fl c_2 = 1 - \left(\frac{19}{10}{\varrho}^{2}-\frac{1}{8}{\varrho}^{4}\right){\psi}^{2} + \frac{11}{15}{\varrho}^{2}{\psi}^{4}, \\
        \fl c_3 = \left[\left(\frac{2711}{3200}-\frac{16}{9\pi^2}\right){\varrho}^{2} + \left(\frac{1027}{2560}-\frac{16}{3\pi^2}\right){\varrho}^{4} - \left(\frac{517}{3072}-\frac{16}{9\pi^2}\right){\varrho}^{6}\right]\psi^{2} \nonumber\\ - \left[\left(\frac{29713}{22400}-\frac{8}{3\pi^2}\right){\varrho}^{2} + \left(\frac{1633}{3840}-\frac{16}{3\pi^2}\right){\varrho}^{4} - \left(\frac{631}{4608}-\frac{16}{9\pi^2}\right){\varrho}^{6}\right]{\psi}^{4} \nonumber\\ + \left(\frac{17749}{33600}-\frac{8}{9\pi^2}\right){\varrho}^{2}{\psi}^{6}.
       \end{eqnarray}
  \item Gravitational mass $M^*$:
        \begin{eqnarray}
         \fl M^*_{\rm NL} = \frac{4 \gamma}{3\pi}, \\
         \fl c_1 = \frac{9}{10} - \frac{1}{15}\,\psi^2, \\
         \fl c_2 = \frac{11}{7}-\frac{64}{9\pi^2} - \left(\frac{58333}{67200}-\frac{8}{\pi^2}\right)\psi^2 + \left(\frac{1429}{33600}-\frac{8}{9\pi^2}\right)\psi^4, \\
         \fl c_3 = \frac{2521}{1260}-\frac{1568}{135 \pi^2} - \left(\frac{393917683}{344064000}-\frac{166013}{15120 \pi^2}\right)\psi^2 - \left(\frac{301426021}{1548288000}-\frac{3175}{3024 \pi^2}\right)\psi^4 \nonumber \\ + \left(\frac{22109}{756000}-\frac{56}{135 \pi^2}\right)\psi^6, \\
         \fl c_4 = \frac{6242}{3465}-\frac{23792}{945 \pi^2}+\frac{4096}{27 \pi^4} - \left(\frac{1725516795817609}{2790386565120000}-\frac{1658106057221}{40874803200 \pi^2}+\frac{1024}{3 \pi^4}\right)\psi^2 \nonumber \\ - \left(\frac{4757113193121521}{9766352977920000}+\frac{2400073720591}{122624409600 \pi^2}-\frac{128}{27 \pi^4}\right)\psi^4 \nonumber \\ - \left(\frac{91538340739}{1300561920000}-\frac{2245701613}{479001600 \pi^2}+\frac{128}{3 \pi^4}\right)\psi^6 \nonumber \\ + \left(\frac{11617705117}{447068160000}-\frac{1907}{3780 \pi^2}+\frac{64}{27 \pi^4}\right)\psi^8.
        \end{eqnarray}
  \item Angular velocity $\Omega^\circ = \Omega M$:
        \begin{eqnarray}
         \fl \Omega^\circ_{\rm NL} = \frac{4 \gamma^{3/2}}{3\pi}\sqrt{1-\psi^2}, \\
         \fl c_1 = \frac{3}{20} + \frac{3}{10}\,\psi^2, \\
         \fl c_2 = \frac{1249}{1120}-\frac{32}{3\pi^2} - \left(\frac{9433}{8960}-\frac{12}{\pi^2}\right)\psi^2 + \left(\frac{14447}{67200}-\frac{4}{3\pi^2}\right)\psi^4, \\
         \fl c_3 = \frac{1219}{1152}-\frac{280}{27 \pi^2} - \left(\frac{18872669}{294912000}-\frac{5681}{4320 \pi^2}\right)\psi^2 - \left(\frac{2768922671}{3096576000}-\frac{8323}{864 \pi^2}\right)\psi^4 \nonumber \\ + \left(\frac{988499}{6048000}-\frac{22}{15 \pi^2}\right)\psi^6, \\
         \fl c_4 = \frac{4861867}{7096320}-\frac{29993}{945 \pi^2}+\frac{6656}{27 \pi^4} \nonumber \\ + \left(\frac{393976735547767}{1860257710080000}+\frac{296409922049}{5449973760 \pi^2} - \frac{1664}{3 \pi^4}\right)\psi^2 \nonumber \\ - \left(\frac{117736707196657}{6510901985280000}+\frac{275650267021}{7431782400 \pi^2}-\frac{10088}{27 \pi^4}\right)\psi^4 \nonumber \\ - \left(\frac{7449051923917}{9537454080000}-\frac{4791322541}{319334400 \pi^2}+\frac{208}{3 \pi^4}\right)\psi^6 \nonumber \\ + \left(\frac{27346973}{197120000}-\frac{53287}{30240 \pi^2}+\frac{104}{27 \pi^4}\right)\psi^8.
        \end{eqnarray}
  \item Angular momentum $J^\circ$:
        \begin{eqnarray}
         \fl J^\circ_{\rm NL} = \frac{3 \pi}{10 \sqrt{\gamma}}\sqrt{1-\psi^2}, \\
         \fl c_1 = -\frac{1}{20} - \frac{1}{14}\,\psi^2, \\
         \fl c_2 = -\frac{77363}{50400}+\frac{416}{27 \pi^2} + \left(\frac{162173}{134400}-\frac{308}{27 \pi^2}\right)\psi^2 - \left(\frac{3211}{86400}-\frac{4}{9 \pi^2}\right)\psi^4, \\
         \fl c_3 = -\frac{361961}{336000}+\frac{488}{45 \pi^2}  + \left(\frac{4379136107}{10597171200}-\frac{429269}{110880 \pi^2}\right)\psi^2  \nonumber \\ + \left(\frac{145460164961}{238436352000}-\frac{683051}{110880 \pi^2}\right)\psi^4 \nonumber \\ - \left(\frac{11652667}{465696000}-\frac{94}{315 \pi^2}\right)\psi^6, \\
         \fl c_4 = \frac{5382534221}{3228825600}+\frac{10753}{4725 \pi^2}-\frac{14848}{81 \pi^4} \nonumber \\ -\left(\frac{619674800505859183}{169283451617280000}-\frac{57472167761}{70849658880 \pi^2} - \frac{28288}{81 \pi^4}\right)\psi^2 \nonumber \\ + \left(\frac{135571135157430463}{84641725808640000}+\frac{3757347355969}{1062744883200 \pi^2}-\frac{15464}{81 \pi^4}\right)\psi^4 \nonumber \\ + \left(\frac{2236889283151}{5904138240000}-\frac{8830544257}{1383782400 \pi^2}+\frac{2096}{81 \pi^4}\right)\psi^6 \nonumber \\ - \left(\frac{206530297}{10567065600}-\frac{135943}{453600 \pi^2}+\frac{8}{9 \pi^4}\right)\psi^8.
        \end{eqnarray}
  \item Magnetic dipole moment $D^\circ$:
        \begin{eqnarray}
         \fl D^\circ_{\rm NL} = \frac{3 \pi}{20 \sqrt{\gamma}}\,\psi\sqrt{1-\psi^2}, \\
         \fl c_1 = \frac{29}{28} + \frac{9}{70}\,\psi^2, \\
         \fl c_2 = -\frac{13099}{5600}+\frac{656}{27 \pi^2} + \left(\frac{1637029}{1209600}-\frac{428}{27 \pi^2}\right)\psi^2 + \left(\frac{9}{86400}+\frac{4}{9 \pi^2}\right)\psi^4, \\
         \fl c_3 = -\frac{14126969}{5174400}+\frac{56044}{2079 \pi^2}  + \left(\frac{79899457229}{43352064000}-\frac{5901503}{332640 \pi^2}\right)\psi^2 \nonumber \\  + \left(\frac{143489392481}{238436352000}-\frac{2100353}{332640 \pi^2}\right)\psi^4 - \left(\frac{326299}{42336000}-\frac{22}{105 \pi^2}\right)\psi^6, \\
         \fl c_4 = \frac{104788527749}{80720640000}+\frac{20606291}{1351350 \pi^2}-\frac{22528}{81 \pi^4}  \nonumber \\ -\left(\frac{2217555015040271693}{507850354851840000}+\frac{2568918307691}{354248294400 \pi^2}-\frac{40768}{81 \pi^4}\right)\psi^2 \nonumber \\ + \left(\frac{45857036770964723}{16928345161728000}-\frac{1399751492287}{1062744883200 \pi^2}-\frac{20744}{81 \pi^4}\right)\psi^4 \nonumber \\ + \left(\frac{16546156571327}{53137244160000}-\frac{25927117379}{4151347200 \pi^2}+\frac{2576}{81 \pi^4}\right)\psi^6 \nonumber \\ - \left(\frac{7617384713}{968647680000}-\frac{29677}{151200 \pi^2}+\frac{8}{9 \pi^4}\right)\psi^8.
        \end{eqnarray}
  \item Gravitational quadrupole moment $m_2^\circ$:
        \begin{eqnarray}
         \fl {m_2^\circ}_{\rm NL} = -\frac{9}{80 \pi^2 \gamma^2}, \\
         \fl c_1 = -\frac{47}{35} - \frac{18}{35}\psi^2, \\
         \fl c_2 = -\frac{4267}{6300}+\frac{464}{27\pi^2} + \left(\frac{3667}{2240}-\frac{512}{27 \pi^2}\right)\psi^2 - \left(\frac{22237}{151200}-\frac{16}{9 \pi^2}\right)\psi^4, \\
         \fl c_3 = \frac{12371}{40425}-\frac{256}{105 \pi^2} - \left(\frac{9772341947}{39739392000}-\frac{50123}{27720 \pi^2}\right)\psi^2 \nonumber \\ + \left(\frac{81384847}{3973939200}+\frac{113}{5544 \pi^2}\right)\psi^4 - \left(\frac{3465793}{58212000}-\frac{64}{105 \pi^2}\right)\psi^6, \\
         \fl c_4 = \frac{11295719}{6006000}-\frac{128}{45 \pi^2}-\frac{4096}{27 \pi^4} \nonumber \\ - \left(\frac{5428030087901639}{1692834516172800}+\frac{7641678347}{2683699200 \pi^2}-\frac{1024}{3 \pi^4}\right)\psi^2 \nonumber \\ + \left(\frac{69407086543177}{46506442752000}+\frac{48426129893}{5904138240 \pi^2}-\frac{6208}{27 \pi^4}\right)\psi^4 \nonumber \\ - \left(\frac{3595050520793}{30996725760000}+\frac{1093286657}{345945600 \pi^2}+\frac{128}{3 \pi^4}\right)\psi^6 \nonumber \\ - \left(\frac{13554494567}{322882560000}-\frac{12287}{18900 \pi^2}+\frac{64}{27 \pi^4}\right)\psi^8.
        \end{eqnarray}
  \item Electric quadrupole moment $e_2^\circ$:
        \begin{eqnarray}
         \fl {e_2^\circ}_{\rm NL} = -\frac{9}{80 \pi^2 \gamma^2}\,\psi, \\
         \fl c_1 = -\frac{59}{35} - \frac{6}{35}\,\psi^2, \\
         \fl c_2 = -\frac{673}{2100}+\frac{464}{27\pi^2} + \left(\frac{78401}{60480}-\frac{512}{27 \pi^2}\right)\psi^2 - \left(\frac{2741}{16800}-\frac{16}{9 \pi^2}\right)\psi^4, \\
         \fl c_3 = \frac{9575749}{9702000}-\frac{10112}{1155 \pi^2} - \left(\frac{23595907079}{17031168000}-\frac{17119}{1320 \pi^2}\right)\psi^2 \nonumber \\ + \left(\frac{1433090333}{2838528000}-\frac{3159}{616 \pi^2}\right)\psi^4 - \left(\frac{10037}{117600}-\frac{32}{35 \pi^2}\right)\psi^6, \\
         \fl c_4 = \frac{16364265889}{10090080000}-\frac{1388}{5005 \pi^2}-\frac{4096}{27 \pi^4} \nonumber \\  - \left(\frac{32617425522286343}{11542053519360000}+\frac{1886067001921}{265686220800 \pi^2}-\frac{1024}{3 \pi^4}\right)\psi^2 \nonumber \\ + \left(\frac{25607911281592511}{21160431452160000}+\frac{436389127799}{37955174400 \pi^2}-\frac{6208}{27 \pi^4}\right)\psi^4 \nonumber \\ + \left(\frac{208667707711}{3444080640000}-\frac{467429993}{94348800 \pi^2}+\frac{128}{3 \pi^4}\right)\psi^6 \nonumber \\ - \left(\frac{1019676659}{17611776000}-\frac{2249}{2700 \pi^2}+\frac{64}{27 \pi^4}\right)\psi^8.
        \end{eqnarray}

\end{itemize}

\end{document}